%% file: main.tex
\documentclass[sigconf]{acmart}
\AtBeginDocument{%
  }

\setcopyright{none}
\copyrightyear{2025}
\acmYear{2025}
\acmDOI{XXXXXXX.XXXXXXX}
\acmConference[Conference acronym 'XX]{Make sure to enter the correct
  conference title from your rights confirmation email}{June 03--05,
  2018}{Woodstock, NY}
\acmISBN{978-1-4503-XXXX-X/2018/06}

\usepackage{array}

\usepackage{float}
\usepackage{pgfplots}
\pgfplotsset{compat=1.18}
\usepackage{xcolor}
\usepackage{tikz}
\usetikzlibrary{calc, positioning}

\definecolor{staticcol}{RGB}{76, 175, 80}
\definecolor{bertcol}{RGB}{244, 67, 54}
\definecolor{qtycol}{RGB}{33, 150, 243}

\pgfplotsset{
    tradeoff style/.style={
        width=4.5cm,
        height=4cm,
        xlabel={\footnotesize Avg Embedding Time (ms/Log)},
        ylabel={\footnotesize Avg F1-score (\%)},
        grid=both,
        grid style={dotted, gray!30},
        tick label style={font=\scriptsize},
        label style={font=\scriptsize},
        title style={font=\small\bfseries, yshift=-4pt},
        enlarge x limits=0.25,
        enlarge y limits=0.15,
    }
}

\definecolor{bertblue}{HTML}{0000FF}   
\definecolor{qtyorange}{HTML}{FFA500}  

\usepackage{dblfloatfix}

\usepackage{algorithm}
\usepackage{algpseudocode}
\usepackage{amsmath}

\usepackage{subcaption}

\usepackage{threeparttable}

\usepackage{tikz}
\usepackage{booktabs,multirow}

\usepackage{threeparttable}

\begin{document}



\title{A Comparative Study of Semantic Log Representations for Software Log-based Anomaly Detection}

\author{Yuqing Wang}
\email{yuqing.wang@helsinki.fi}
\affiliation{%
  \institution{University of Helsinki}
  \city{Helsinki}
  \country{Finland}
}

\author{Ying Song}
\email{ying.song@helsinki.fi}
\affiliation{%
  \institution{University of Helsinki}
  \city{Helsinki}
  \country{Finland}
}

\author{Xiaozhou Li}
\email{Xiaozhou.Li@unibz.it}
\affiliation{%
  \institution{Free University of Bozen-Bolzano}
  \city{Bolzano}
  \country{Italy}}

\author{Nana Reinikainen}
\affiliation{%
  \institution{University of Helsinki}
  \city{Helsinki}
  \country{Finland}}
\email{nana.reinikainen@helsinki.fi}

\author{Mika V. M{\"a}ntyl{\"a}}
\affiliation{%
  \institution{University of Helsinki}
  \city{Helsinki}
  \country{Finland}}
\email{mika.mantyla@helsinki.fi}

\renewcommand{\shortauthors}{Wang et al.}

\begin{abstract}




Recent deep learning (DL) methods for log anomaly detection increasingly rely on semantic log representation methods that convert the textual content of log events into vector embeddings as input to DL models. However, these DL methods are typically evaluated as end-to-end pipelines, while the impact of different semantic representation methods is not well understood. 

In this paper, we benchmark widely used semantic log representation methods, including static word embedding methods (Word2Vec, GloVe, and FastText) and the BERT-based contextual embedding method, across diverse DL models for log-event level anomaly detection on three publicly available log datasets: BGL, Thunderbird, and Spirit. We identify an effectiveness–efficiency trade-off under CPU-only deployment settings: the BERT-based method is more effective, but incurs substantially longer log embedding generation time, limiting its practicality; static word embedding methods are efficient but are generally less effective and may yield insufficient detection performance. 

Motivated by this finding, we propose QTyBERT, a novel semantic log representation method that better balances this trade-off. QTyBERT uses SysBE, a lightweight BERT variant with system-specific quantization, to efficiently encode log events into vector embeddings on CPUs, and leverages CroSysEh to enhance the semantic expressiveness of these log embeddings. CroSysEh is trained unsupervisedly using unlabeled logs from multiple systems to capture the underlying semantic structure of the standard BERT model’s embedding space. We evaluate QTyBERT against existing semantic log representation methods. Our results show that, for the DL models, using QTyBERT-generated log embeddings achieves detection effectiveness comparable to or better than BERT-generated log embeddings, while bringing log embedding generation time closer to that of static word embedding methods.

\end{abstract}



\acmJournal{PACMSE}
\received{20 February 2007}
\received[revised]{12 March 2009}
\received[accepted]{5 June 2009}

\maketitle

\input{sections/introduction}

\input{sections/background}
\input{sections/preliminaryStudy}

\input{sections/method}

\input{sections/experiment}

\input{sections/results}

\input{sections/conclution}

\bibliographystyle{ACM-Reference-Format}
\bibliography{mainbib}


\end{document}

%% file: sections/introduction.tex
\section{Introduction}
As modern software systems become increasingly complex, the potential for anomalies grows \cite{LogRobust}. The anomalies may arise from various causes, e.g., 
misconfigurations, resource contention, or unpredictable workloads \cite{Adha2025}.  Even a small anomaly may compromise system reliability and performance 
\cite{le2022log}.  Timely and effective anomaly detection is critical to prevent anomalies from escalating into severe failures 
\cite{Adha2025,CroSysLog}. Software logs record runtime information and system states, providing a primary source for
anomaly detection \cite{LogRobust}.  However, modern systems generate logs at a massive scale. Recent reports indicate that many systems produce more than 1 TB of logs per day \cite{peronto2024logtrends}. This makes manual log anomaly detection labor-intensive and error-prone.


Deep learning (DL) methods have been widely adopted for automated log anomaly detection. A critical step in these methods is \textit{log representation}, which converts log events into structured inputs
for DL models \cite{Adha2025}. Semantic log representation methods have been increasingly adopted in recent DL studies \cite{LiHeng2023,LogRobust}. Compared to traditional methods that represent logs using discrete features (e.g., event identifiers or occurrence counts), \textit{semantic log representation} methods encode the textual content of log events into vector embeddings that preserve their semantic meaning, thus providing more informative inputs for DL models \cite{LiHeng2023,Adha2025}. Several semantic log representation methods have been proposed, ranging from methods based on static word embedding models (e.g., FastText) to pre-trained language models (e.g., BERT). DL methods built on such representations have shown promising effectiveness across diverse real-world log datasets \cite{Adha2025,NeuralLog2021}. However, in prior work, these methods are typically evaluated as end-to-end pipelines that couple semantic representations with DL models, making the reported performance reflect the overall pipeline \cite{LiHeng2023}. 
It remains unclear how different semantic log representation methods affect the performance of the DL methods.

 
To address the gap, we conduct a comprehensive empirical study to evaluate four widely used semantic log representation methods, including static word embedding methods (Word2Vec, GloVe,
and FastText) and the BERT-based contextual embedding method, across a broad set of DL models (covering popular recurrent, convolutional, and attention-based architectures) using publicly available log datasets from three large-scale distributed systems: BGL, Thunderbird (TB), and Spirit. We focus on log event-level anomaly detection, which is well-suited for such distributed software systems where log events are generated in an interleaved manner by different system components that operate independently or participate in inter-component interactions; it enables fine-grained anomaly localization by identifying
the responsible components, thereby facilitating root cause analysis \cite{CroSysLog,hashemi2024onelog}. This setting differs from log session-level anomaly detection, which determines whether a session of log events is anomalous or normal. In our datasets, explicit session boundaries are not provided, and constructing sessions would require system-specific heuristics that may introduce confounding factors for our evaluation. We examine both detection effectiveness and computational efficiency, with a particular focus on CPU-only deployment settings that are common in production environments. Although DL models are typically trained using GPUs, not all production environments have dedicated GPU resources; also, log anomaly detection needs to process log events continuously, and provisioning GPUs for sustained inference can significantly increase operational costs \cite{EdgeLog,LightLog,Adha2025}.





The results of our empirical study reveal a clear effectiveness-efficiency trade-off under CPU-only deployment settings: the
BERT-based method is more effective, but incurs substantially longer log embedding generation time, limiting its practicality; static word embedding methods are efficient, but are generally less effective and may yield insufficient detection performance.


Motivated by this finding, we propose QTyBERT, a novel semantic log representation method that better balances this effectiveness–efficiency trade-off in CPU deployment settings. 
The key idea behind QTyBERT is to use a lightweight BERT variant to efficiently generate log embeddings while ensuring that these embeddings achieve semantic expressiveness comparable to that of the standard BERT model. Although lightweight variants of BERT have been widely explored in the natural language processing (NLP) community as efficient alternatives for BERT-style contextual embedding generation \cite{tinybert,distilbert}, their applicability to semantic log representation remains unexplored.

QTyBERT consists of two components: a System-specific Base Encoder (SysBE), which converts log events into vector embeddings, and a Cross-System Embedding Enhancement module (CroSysEh), which operates on these log embeddings to improve their semantic expressiveness. SysBE is constructed by applying system-specific quantization to a lightweight BERT variant, enabling efficient log embedding generation on CPUs. CroSysEh is trained
in an unsupervised manner using unlabeled logs from multiple systems to capture the underlying semantic structure of the standard BERT model’s embedding
space, compensating for the semantic loss introduced by the compact design of SysBE.

We evaluate QTyBERT against existing semantic log representation methods under the same experimental settings as our empirical study. 
Our results show that, for the same DL models, QTyBERT-generated log embeddings achieves detection
effectiveness comparable to or better than BERT-generated log embeddings, with F1-score differences within $\pm$1\% in most cases and improvements of up to 21.53\%. QTyBERT reduces log embedding generation time by more than 94\% compared to BERT, achieving sub-millisecond latency per log event, bringing its efficiency much closer to that of static word embedding methods. 

In summary, our main contributions are highlighted as follows:
\begin{itemize}
    \item We conduct a comprehensive empirical study to benchmark widely used semantic log representation methods across a broad set of DL models for log event-level anomaly detection using publicly available log datasets.
    
    \item Our empirical study identifies a clear trade-off between static word embedding and BERT-based methods in detection effectiveness and log embedding generation efficiency under CPU-only deployment settings.
    
    \item We propose QTyBERT, a novel semantic log representation method that better balances this effectiveness–efficiency trade-off, and evaluate it against existing semantic log representation methods using publicly available log datasets.
    
    
    

\end{itemize}

%% file: sections/background.tex
\section{Background}
\subsection{Deep Learning-based Log Anomaly Detection with Semantic Log Representation}
\subsubsection{Semantic Log Representation.} 
\label{sec:EmS_LogRpre}
Early studies use static word embedding methods, which first parse log messages into structured log templates, decompose these templates into word tokens, and then encode word tokens into vector representations using pre-trained static embedding models. The widely used static embedding models include Word2Vec \cite{Word2Vec}, GloVe \cite{glove}, and FastText \cite{Fasttext}. For instance, Word2Vec is used in TinyLog \cite{TinyLog}, LightLog \cite{LightLog}, and EdgeLog \cite{EdgeLog}, where a Word2Vec model is either trained on the target system log templates or pre-trained and then applied to encode each template into vector embeddings. 
GloVe is adopted in LogTransfer \cite{LogTransfer} and LogPal \cite{LogPal}, while FastText is employed in LogRobust \cite{LogRobust} and RT-Log \cite{RT-Log}, both using 
pre-trained word embeddings trained on large-scale corpora (e.g., Common Crawl) to encode word tokens of log templates into vector embeddings.  

Static word embedding methods are computationally efficient because they use pre-trained word embeddings to represent words in log templates. Generating such log embeddings mainly involves dictionary lookup and aggregation operations, making them suitable for resource-constrained environments such as CPU-only deployments \cite{Word2Vec,Fasttext,glove}. However, these methods have two key limitations. First, they rely on a fixed word vocabulary learned from training data and thus struggle to handle out-of-vocabulary (OOV) words, which are common in software logs \cite{NeuralLog2021,LEE2023}. Examples of OOV words include system modules (e.g., `kubelet', `etcd'), kernel-related processes (e.g., `ksoftirqd', `rcu\_sched') \cite{CroSysLog}. Second, static word embedding methods depend on log parsing, which separates static (template) and variable (parameters) part of the log, e.g., in log message ``User connected to 192.168.0.1'' the template is ``User connected to" and parameter is ``192.168.0.1''.  The quality of log embeddings from these methods depends on the accuracy of log parsing, which can affect the effectiveness of downstream anomaly detection \cite{NeuralLog2021,LEE2023}. Even widely used parsers such as Drain \cite{Drain} may produce parsing errors due to inconsistent log formats, nested data structures, or missing values \cite{Sedki2023,ParsingError2023}.

Recent studies have shifted to the  BERT-based method for contextual embedding (e.g., NeuralLog \cite{NeuralLog2021}, CNN \cite{Baseline1dCNN}, CroSysLog \cite{CroSysLog}), which use the pre-trained language model BERT to encode raw log events into vector representations.
This BERT-based method addresses the limitations of static word embedding methods. It does not require log parsing and can directly process raw log events. It first tokenizes each log event into subword tokens and then encodes these subword tokens using BERT's self-attention mechanism, capturing semantic relationships among subword tokens. This mechanism allows to handle OOV words by decomposing them into subwords. With this mechanism, the embedding of each subword token is contextualized, i.e., it is dynamically generated based on its surrounding subword tokens. However, generating such BERT-based log embeddings is computationally expensive. In practice, BERT inference for embedding generation is typically accelerated using GPUs \cite{devlin-etal-2019-bert, suppa-etal-2021-cost, LEE2023}.

\subsubsection{Deep Learning Models.} \label{sec:relatedWorkDL} Software system logs are  sequential, as log events are generated over time during system execution \cite{Adha2025,LogRobust}. Log events exhibit temporal correlations that reflect system behavior. DL-based sequence models are therefore widely adopted to capture such temporal dependencies. Recurrent neural network (RNN) variants are the most commonly used DL models. For example, CroSysLog \cite{CroSysLog} and LogAnomaly \cite{LogAnomaly} use LSTM, while the study \cite{GRUBaseline} uses GRU. SwissLog \cite{SwissLog} and LogRobust \cite{LogRobust} use the Attention-based BiLSTM (AttBiLSTM), which extends LSTM with a bidirectional encoder and an attention mechanism to capture both forward and backward dependencies among log events and focus on the most relevant ones. NeuralLog \cite{NeuralLog2021} and HitAnomaly \cite{HitAnomaly} use a Transformer-encoder (TransEnc), which replaces recurrence with self-attention to capture long-range dependencies across log events. CNN has also been applied, e.g., in the studies \cite{Baseline1dCNN,LiqiangCNN, hashemi2024onelog}. Unlike RNN models, CNNs apply convolutional filters over log event sequences to capture local patterns among neighboring log events.

\subsection{Related Work}

\subsubsection{Effect of Semantic Log Representation Methods} \label{sec:relatedWordSemantic}  The studies on how semantic log representation methods affect DL-based log anomaly detection are scarce. The only closely related work is by Wu et al. \cite{LiHeng2023}, who investigate the impact of log representation methods on log session-level anomaly detection. 
Their results show that, 
BERT-generated log embeddings achieve the highest effectiveness when used with DL models, while classical log representation methods such as MCV outperform semantic-based ones when used with traditional ML models. Our empirical study addresses several important aspects not considered in their study. First, our study investigates log event-level anomaly detection, which is not explored in their study. Second, their study evaluates three semantic log representation methods (Word2Vec, FastText, and BERT), whereas our study additionally includes GloVe, which is also widely used in existing DL-based log anomaly detection. Third, we evaluate each log representation method on a broader set of DL models, covering commonly used ones, including RNN, GRU, LSTM, AttBiLSTM, TransEnc, and CNN, whereas their study only covers MLP, CNN and LSTM. Last, we benchmark the computational efficiency of each semantic log representation method, which is a crucial practical concern when deploying these methods in production environments but was not previously evaluated.

\subsubsection{Efficient BERT-style Log Embedding Generation.} Efforts to address the high computational cost of BERT-based log embedding generation remain limited. The only related work is LAnoBERT \cite{LEE2023}, which introduces a log dictionary-based inference mechanism to avoid redundant embedding computation for previously seen log events, but the computational cost of generating embeddings for new log events remains high.

In the NLP community, lightweight variants of BERT, such as DistilBERT \cite{distilbert} and TinyBERT~\cite{tinybert}, have been proposed to accelerate BERT-style contextual embedding generation in resource-constrained environments. These variants compress the standard BERT model using techniques such as knowledge distillation and architectural compression, resulting in fewer model layers and parameters and thus reducing computational cost during embedding generation \cite{Bertcompress}. However, this efficiency comes at the cost of semantic loss, as their ability to capture complex semantic and contextual relationships among subword tokens is weakened compared to the standard BERT \cite{Bertcompress,tinybert}. As such, applying these variants to domain-specific tasks typically requires fine-tuning~\cite{tinybert,distilbert}, which involves task-specific training with domain data and updating model parameters. This process incurs additional training costs and must be repeated for each task. These variants have been widely adopted as efficient alternatives to the standard BERT for NLP tasks, e.g., text classification and question answering \cite{tinybert}. However, their applicability to semantic log representation in log anomaly detection remains unexplored. This motivates us to develop QTyBERT.

Our QTyBERT addresses the gaps from two aspects. First, inspired by lightweight BERT variants in NLP, QTyBERT extends this idea to efficient log embedding generation through SysBE, a lightweight BERT variant with system-specific quantization. Unlike LAnoBERT, SysBE directly accelerates log embedding generation on CPUs. Second, to compensate for the semantic loss introduced
by the compact design of SysBE, QTyBERT employs CroSysEh, which operates on log embeddings generated by SysBE to improve their semantic expressiveness, eliminating per-system fine-tuning and reducing such training costs in multi-system settings.

%% file: sections/preliminaryStudy.tex
\section{Empirical Study}

Our empirical study is guided by the research question:

\begin{itemize}
    \item RQ1: How do different semantic log representation methods impact the effectiveness and efficiency of log event-level anomaly detection, when serving input for DL models? 
\end{itemize}

\subsection{Experimental Setup}
\label{sec:EmS_emperiment_setup}
\subsubsection{Datasets}
For a comprehensive evaluation, we use software log datasets of three large-scale distributed supercomputing systems: BGL, TB, and Spirit, sourced from the USENIX CFDR repository \cite{USENIX_CFDR,Oliner2007}. BGL is the IBM Blue Gene/L system at Lawrence Livermore National Laboratory. TB and Spirit are high-performance Linux clusters operated by Sandia National Laboratories. Each dataset includes log event level binary labels (normal vs. anomalous). We used two chronological log sequences from each system: one sequence as the training set, and the other as the testing set. Table \ref{tab:ano_datasets_slim} summarizes the statistics of these sets for each system. For each system, the testing set is temporally subsequent to the training set to preserve chronological order;  these two sets do not overlap, there is a temporal gap of 4-6 months between the training set and the testing set to break short-range autocorrelation and avoid near-duplicate patterns around the boundary between the two sets, thereby improving the validity of the evaluation \cite{Cerqueira2020,hespeler2025}. 

\begin{table}[ht]
\centering
\caption{Statistics of training and testing sets.}
\label{tab:ano_datasets_slim}
\setlength{\tabcolsep}{6pt}
\begin{tabular}{llrr}
\toprule
System & Set & \# Log events & \# Anomaly \\
\midrule
\multirow{2}{*}{BGL} 
    & Training & 1,885,397 & 227,994 (12.09\%) \\
    & Testing  &   471,349 & 37,000 (7.85\%)  \\ 
\midrule
\multirow{2}{*}{TB} 
    & Training & 997,677 & 69,838 (7.00\%) \\
    & Testing  & 1,396,747 & 184,231 (13.19\%)  \\ 
\midrule
\multirow{2}{*}{Spirit} 
    & Training & 499,095 & 149,728 (30.00\%) \\
    & Testing  & 499,095 & 19,964 (4.00\%)  \\ 
\bottomrule
\end{tabular}
\end{table}

\subsubsection{Pre-processing} \label{sec:EmbPreProcess} For each system, we utilize LogLead \cite{LogLead} to process raw log files, extracting individual log events and organizing them into dataframes that capture key attributes such as timestamp, severity level, reporting component, log message, and anomaly label, if these are available. Since the attributes vary across datasets, we remove log events with missing values in the attributes defined by each dataset, and then sort the remaining log events in chronological order to reflect the operational sequence. For each log event, we concatenate the textual attributes (i.e., reporting component, severity level, and log message) into a single text sequence to represent this log event. The concatenated sequence is then preprocessed by lowercasing, removing non-alphabetic characters, and masking sensitive variables, e.g., replacing ``192.168.1.*" with ``ip address",  or ``/var/app/config/settings.yaml'' to ``file path''. This design differs from conventional log session-level anomaly detection, where log messages alone are used to represent log events, as anomaly signals are typically captured from patterns in log sequences \cite{Adha2025}. Since we focus on log event-level anomaly detection in distributed systems, where log events are generated by different system components that operate independently or participate in inter-component interactions, each log event is expected to carry sufficient information for anomaly detection. Therefore, we retain additional textual fields such as reporting component and severity level to preserve component-level operational context, consistent with prior work on this topic \cite{CroSysLog,NeuralLog2021}.


\subsubsection{Semantic Log Representation Methods} \label{sec:EmStudyLogRepreMethod} We evaluate four widely used semantic log representation methods: three static word embedding methods (Word2Vec, FastText, and GloVe), and the BERT-based contextual embedding method. We follow prior studies reviewed in Section \ref{sec:EmS_LogRpre} to implement these methods and ensure a fair comparison across them. For the static word embedding methods, we use the pre-trained FastText model (300-dimensional, trained on Common Crawl) \cite{Fasttext}, pre-trained GloVe model (300-dimensional, trained on Wikipedia and Gigaword) \cite{glove}, and train Word2Vec on each system’s training set. For each static word embedding method, we obtain log event-level embeddings as follows: we first parse log events into log templates using Drain \cite{Drain}, tokenize log templates into word tokens, obtain word token embeddings using the corresponding static word embedding model, and then aggregate the token embeddings using TF-IDF weighting. We keep the log parser, tokenization strategy, and aggregation approach fixed across these methods to avoid introducing confounding factors that affect our evaluation results. For the BERT-based method, we implement it using a neural representation approach following prior studies \cite{NeuralLog2021, CroSysLog,Baseline1dCNN}. Specifically, we use the BERT-base model \cite{bert2018}, which consists of 12 Transformer encoder layers with 768 hidden units and 12 attention heads. We obtain log event-level embeddings as follows: we tokenize log events into subword tokens using WordPiece technique \cite{wu2016google}, feed subword tokens into BERT-base to obtain contextualized subword embeddings, and then aggregate these subword embeddings using mean pooling over the final hidden layer.



\subsubsection{Deep Learning Models} \label{EmS_DLmodels} We select commonly used DL models in prior log anomaly detection studies, including all discussed in Section \ref{sec:relatedWorkDL}: GRU, LSTM, AttBiLSTM, CNN, and TransEnc. In addition, we include a vanilla RNN as a simple recurrent baseline to assess the benefits of more complex recurrent architectures. For each system, we train all DL models on its training set using a consistent supervised setting and evaluate them on its testing set. This ensures that the comparison focuses solely on the effect of different log representation methods, rather than differences caused by unsupervised detection objectives or thresholding strategies. These DL models use log embeddings generated by each log representation method (Section \ref{sec:EmStudyLogRepreMethod}) as input during both training and testing. Following prior work on log event-level anomaly detection \cite{CroSysLog}, these DL models take fixed-size windows of log event embeddings as inputs. Specifically, for each system $s_j$, its log events are ordered chronologically as $L^{(j)} = \{e_1, e_2, \ldots, e_N\}$, where each $e_k$ denotes the embedding of the $k$-th log event produced by a certain log representation method. We partition $L^{(j)}$ into non-overlapping windows of size $m$, where each window consists of $m$ consecutive log event embeddings and serves as an input to the DL models.

\subsubsection{Implementation Details} \label{EmS_imple} We perform the model training on a computing server with 16 CPU cores and a single NVIDIA Ampere A100 GPU with 40 GB of memory. All DL models are trained for a fixed number of epochs, and each model is tuned to obtain its optimal performance under our experimental setting. During testing,  we simulate CPU-only environments with 4-core or 8-core CPU allocations without GPU resources. These environments are configured to ensure full utilization of the allocated CPU cores. We monitor CPU utilization throughout the evaluation process.

\subsubsection{Metrics.} \label{EmS_metrics} For each DL model, we compare different semantic log representation methods in terms of anomaly detection effectiveness using Precision, Recall, and F1-score. These metrics are computed based on True Positives (TP), False Positives (FP), and False Negatives (FN). Precision is defined as the proportion of correctly identified anomalies among all predicted anomalies, i.e., Precision = \(\frac{TP}{TP + FP}\). Recall measures how many actual anomalies were correctly detected, i.e., Recall = \(\frac{TP}{TP + FN}\). F1-score, as the harmonic mean of Precision and Recall, is given by F1-score = \(\frac{2 \cdot \text{Precision} \cdot \text{Recall}}{\text{Precision} + \text{Recall}}\). We adopt these metrics because log anomaly detection is a binary classification task where the normal and abnormal classes are often imbalanced. In such cases,  Precision quantifies the false alarm rate, Recall ensures that actual anomalies are not missed, and the F1-score offers a balanced summary of both. For efficiency, we compare each log representation method in terms of the time required to generate embeddings for log events in the testing set of each system, as well as the detection latency of each DL model using these log embeddings.

\subsection{Study results and analysis}
\subsubsection{Effectiveness.} The DL models consistently achieve higher effectiveness when using BERT-based log embeddings than those generated by static word embedding methods, with the impact being the most pronounced on BGL. As shown in Table \ref{tab:study_effet_results}, static word embedding methods only achieve F1-scores of 55.05\%--67.87\% on BGL across all DL models; however, replacing them with BERT-based log embeddings improves F1-scores by approximately 13\%--31\% for each model. On TB and Spirit, BERT-based log embeddings remain more effective in most cases, although the performance gap becomes smaller, generally within 9\% F1-score across DL models. These findings are consistent with Wu et al. \cite{LiHeng2023}, who observe similar results for log session-level anomaly detection. In contrast, the performance differences among static embedding methods are limited. Using FastText-, GloVe-, and Word2Vec-based log embeddings, the maximum F1-score deviation on each DL model is small (typically within about 3\%), indicating that the choice among static embedding methods has only a limited impact. 

\subsubsection{Efficiency.} 
\label{sec:EmS_efficiency}
\textbf{Log Embedding Generation.} Static word embedding methods require substantially less log embedding generation time than the BERT-based method under CPU-only environments. As reported in Table \ref{tab:log_embedding_generation_time}, Word2Vec is the fastest across all systems. Under the 8-core CPU setting, Word2Vec requires only 0.05–0.12 ms per log event, whereas BERT requires 4.38–7.44 ms, resulting in approximately 37×–149× longer embedding generation time for BERT. Under the 4-core CPU setting, this gap further widens to approximately 74×–312×.

\textbf{Detection latency.} Compared with log embedding generation time, downstream detection latency is much less affected by the choice of semantic log representation methods. Since DL models using FastText-, GloVe-, and Word2Vec-based log embeddings exhibit very similar detection latency, we report their average latency (Static Avg) along with the maximum deviation ($\Delta_{\max}$) to simplify comparison in Table \ref{tab:detection_latency}. For most DL models (LSTM, GRU, CNN, and RNN), using BERT-based log embeddings incurs only approximately 1.05×–1.20× the detection latency compared to those produced by static word embedding methods. The gap becomes more noticeable for DL models with more complex architecture (TransEnc and AttnBiLSTM), where the increase ranges from approximately 1.13× to 1.9×, depending on the system and CPU configuration. This difference mainly stems from variations in embedding dimensionality across log representation methods: 768 for BERT vs. 300 for static word embedding methods, see Section \ref{sec:EmStudyLogRepreMethod}. Since the computational cost of linear transformations and attention mechanisms scales with the input dimensionality, the higher-dimensional BERT-based log embeddings result in increased processing time in DL models. 


\subsubsection{Trade-off} Our above results show that the choice of semantic log representation methods affects the performance of DL-based log  event-level anomaly detection. BERT-based and static word embedding methods exhibit a clear trade-off between detection effectiveness and log embedding generation efficiency. 
BERT-based log embeddings generally lead to higher detection effectiveness, but their substantially higher generation time may limit their practicality in CPU-only environments. In contrast, static word embedding methods are efficient and well-suited for CPU-only deployment settings, but their log embeddings are generally less effective and may yield insufficient detection performances.

%% file: sections/method.tex
\section{QTyBERT for semantic log representation}




\subsection{Design} 
Figure 1 shows the overall workflow of QTyBERT. During application in a target system, SysBE produces log embeddings, which are then processed by CroSysEh to obtain the final log representations.  We explain how each component is built in the following subsections. 


\begin{figure}[ht]  
  \centering
  \includegraphics[width=1\linewidth]{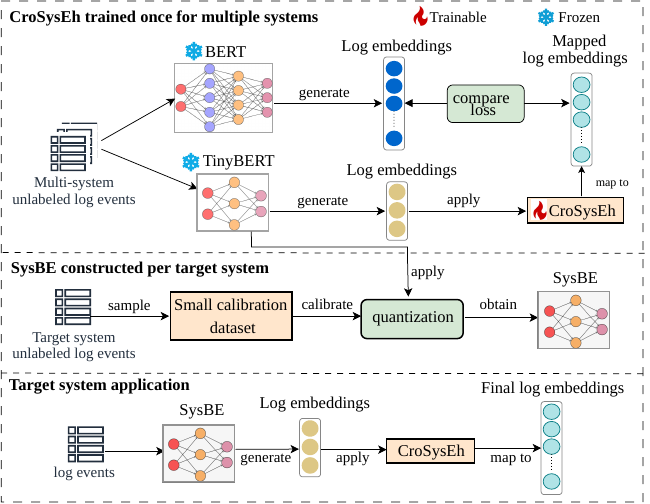} 
   \vspace{-0.7cm}
  \caption{An overview of QTyBERT}
  \label{fig:Overview}
  \Description{Our framework overview}
\end{figure}

\subsubsection{SysBE}
To build SysBE, we conduct a preliminary study on existing lightweight BERT variants for efficient BERT-style contextual embedding generation in CPU-only
environments. Through a review of relevant studies and publicly available implementations, we identify several candidate models (e.g., TinyBERT \cite{tinybert}, DistilBERT \cite{distilbert}, MiniLM \cite{MiniLM}) that retain the standard BERT embedding pipeline, particularly subword tokenization and contextualized subword representations, thereby enabling fair comparison and avoiding the introduction of confounding factors. We evaluate these models under the same experimental settings as in our empirical study. TinyBERT achieves the best effectiveness among the candidates across DL models, while exhibiting comparable embedding generation latency. We therefore select TinyBERT to build SysBE. Specifically, we use a TinyBERT model consisting of 4 Transformer encoder layers with 312 hidden units and 12 attention heads.



Let $\mathcal{M}$ denote the original TinyBERT. For each system $s_j$, we quantize $\mathcal{M}$ to obtain its SysBE in several steps. First, we use a small number of unlabeled log events from $s_j$ to build a calibration dataset $\mathcal{D}_{\text{cal}}^{(j)}$. These log events are not required to be temporally consecutive. They are preprocessed using the same steps as those in our empirical study (Section \ref{sec:EmbPreProcess}). Second, we collect statistics (including value ranges, means, variances, and outliers) from the activations of $\mathcal{M}$ when using $\mathcal{M}$ to generate log embeddings for log events in $\mathcal{D}_{\text{cal}}^{(j)}$, and then use these statistics to calibrate the quantization parameters. Third, based on the calibrated parameters, we quantize approximately 20\% of the linear layers in the Transformer encoders of  $\mathcal{M}$ by mapping their FP32 weights to INT8 representations. Here, FP32 and INT8 denote 32-bit floating-point and 8-bit integer numerical representations, respectively. Our quantization keeps $\mathcal{M}$'s embedding layers and activations in FP32 to maintain 
semantic fidelity, as we empirically observe that aggressive quantization of these components degrades embedding quality, manifested by reduced anomaly detection effectiveness of downstream DL models when operating on the resulting log embeddings. This observation is consistent with prior work that examines how quantize different components of BERT affects the quality of embeddings \cite{static_quantizatio}. We thus obtain the quantized TinyBERT model as the SysBE for $s_j$, denoted as $\mathcal{M}_{\text{q}}^{(j)}$. We export $\mathcal{M}_{\text{q}}^{(j)}$ as an ONNX computation graph \cite{onnx-intro}. The graph includes tensor-level quantization and dequantization operators configured for INT8 precision, which serve as precision bridges between INT8 and FP32 and enable mixed-precision execution.

\subsubsection{CroSysEh}
\label{sec:emb_distillation}
We train CroSysEh using unlabeled log events from multiple systems. We consider $N$ software systems, each producing log events in chronological order. From each system, we randomly sample $m$ unlabeled log events, which are not required to be consecutive.
The sampled log events from all systems constitute a cross-system training dataset, denoted as $\mathcal{D}_{\text{cro}} = {x_1, x_2, \dots, x_n}$, where $x_i$ is each log event.  We pre-process log events in $\mathcal{D}_{\text{cro}}$ using the same steps as those in our empirical study (Section \ref{sec:EmbPreProcess})

We train CroSysEh in several steps, as outlined in Algorithm~\ref{algori:distallation}. 
For each log event $x_i$ from $\mathcal{D}_\text{cro}$, we use both the frozen standard BERT and the frozen original TinyBERT $\mathcal{M}$ to generate the corresponding log embeddings, following the same BERT-based neural representation approach as in our empirical study (Section \ref{sec:EmStudyLogRepreMethod}). We use BERT-base \cite{bert2018} as the standard BERT implementation, consistent with our empirical study setting (Section \ref{sec:EmStudyLogRepreMethod}). For each log event $x_i$, we denote its embedding from BERT as the teacher embedding $h_T \ in \ d_T
$, and the one from $\mathcal{M}$ as the student embedding $h_S \ in \ d_S 
$, where the embedding dimensions correspond to the hidden sizes of each model. We use a residual low-rank function to map $h_S$ to the embedding space of $h_T$: 
\[
h'_S \leftarrow \phi (h_S) = h_S + B(A(h_S))
 \]
 \noindent 
where \( A \in \mathbb{R}^{r \times d_S} \) and \( B \in \mathbb{R}^{d_T \times r} \) are trainable projection matrices, and $r$ is a small bottleneck dimension 
that controls the adaptation capacity. The matrices $A$ and $B$ together parameterize CroSysEh, denoted by $\phi$, which maps each student embedding $h_S$ to the embedding space of $h_T$. We train CroSysEh $\phi$ by minimizing the mean squared error (MSE) between the mapped embedding $h'_S = \phi(h_S)$ and the teacher embedding $h_T$ for each log event $x_i$ in $\mathcal{D}_\text{cro}$. 
The loss function is defined as:
\[
\mathcal{L} = \frac{1}{|\mathcal{D}_{\text{cro}}|} \sum_{x_i \in \mathcal{D}_{\text{cro}}} \left\| h'_S - h_T\right\|_2^2
\]

\noindent During training, we keep both BERT and $\mathcal{M}$ frozen, and optimize only CroSysEh $\phi$ by minimizing the loss $\mathcal{L}$ using gradient descent:
\[
\phi \leftarrow \phi - \eta \cdot \nabla_\phi \mathcal{L}
\]

\noindent where $\eta$ is the learning rate, and $\nabla_\phi \mathcal{L}$ denotes the gradient of $\mathcal{L}$ with respect to the parameters of $\phi$. After training, we obtain the optimized CroSysEh $\phi'$, which maps $\mathcal{M}$’s log embeddings to the embedding space of BERT. Depending on the source of the sampled log events, $\phi'$ can be shared 
across systems.

\begin{algorithm}
\caption{CroSysEh training}
\label{algori:distallation}
\begin{algorithmic}[1]
\Require Log dataset $\mathcal{D}_{\text{cro}} = \{x_1, \ldots, x_n\}$, frozen \textit{BERT}, frozen $\mathcal{M}$, trainable CroSysEh $\phi$, learning rate $\eta$, number of epochs $E$
\State Initialize $\phi$ (i.e., projection matrices $A$ and $B$) randomly
\For{epoch = $1$ to $E$}
    \For{ $x_i$ in $\mathcal{D}_{\text{cro}}$}
        \State $h_T \gets \text{BERT}(x_i)$ 
        \State $h_S \gets \text{$\mathcal{M}$}(x_i)$ 
        \State $h'_S \gets \phi(h_S) = h_S + B(A(h_S))$ 
        \State $\mathcal{L} \gets \mathcal{L} + \|h'_S - h_T\|_2^2$
    \EndFor
    
     \State $\mathcal{L} \gets \mathcal{L} / |\mathcal{D}_{\text{cro}}|$
    \State Update $\phi \gets \phi - \eta \nabla_{\phi} \mathcal{L}$
\EndFor
\State \Return Optimized CroSysEh $\phi'$
\end{algorithmic}
\end{algorithm}

%% file: sections/experiment.tex
\subsection{Experiment setup}
\begin{table*}[!ht]
\caption{Precision, Recall, and F1-score of deep learning models using different semantic log representation methods.}
\label{tab:study_effet_results}
\centering
\begin{tabular}{l p{1.1cm}p{1.1cm}p{1.2cm} p{1.1cm}p{1.1cm}p{1.2cm} p{1.1cm}p{1.1cm}p{1.2cm}}
\toprule
DL Model [Log Rep. Method]
& \multicolumn{3}{c}{BGL}
& \multicolumn{3}{c}{TB}
& \multicolumn{3}{c}{Spirit} \\
\cmidrule(lr){2-4} \cmidrule(lr){5-7} \cmidrule(lr){8-10}
& Precision & Recall & F1-score & Precision & Recall & F1-score & Precision & Recall & F1-score \\ \hline

TransEnc [FastText] 
& 63.55 & 64.47 & 64.01 & 99.91 & 94.56 & 97.16 & 98.73 & 77.22 & 86.67 \\

TransEnc [Glove] & 67.75 & 66.07& 66.90 & 99.99 & 94.54 & 97.19 & 99.34 & 74.75 &85.31\\ 

TransEnc [Word2Vec] & 69.81 &66.04 & 67.87 & 99.41 & 94.55 & 96.92 & 97.40& 74.26& 84.26\\
TransEnc [BERT]
& 92.63 & 88.73 & 90.63
& 90.26 & 94.43 & 92.29
& 98.19 & 80.69 & 88.58 \\

TransEnc [QTyBERT] & 93.46 & 86.76 & 89.98 & 91.17 & 93.14 & 92.15 & 95.51 & 84.16 & 89.47 \\ \hline

AttBiLSTM [FastText]  
& 72.97 & 56.59 & 63.74 & 99.79 & 89.04 & 94.11 & 96.97 & 79.21 & 87.19 \\

AttBiLSTM [Glove] & 72.91 & 59.21 & 65.35 & 99.97 & 85.19 & 91.99 & 98.73 & 77.22 & 86.66 \\

AttBiLSTM [Word2Vec] & 68.43 & 57.42 & 62.45 & 99.99 & 87.38& 93.26 &97.47 & 76.24 & 85.56 \\
AttBiLSTM [BERT]
& 93.33 & 80.43 & 86.40
& 99.00 & 93.80 & 96.32
& 100.0 & 82.67 & 90.51 \\

AttBiLSTM [QTyBERT] & 93.35 & 90.23  & 91.77 & 99.33 & 94.03 & 96.06 & 99.42 &84.65 & 91.44 \\ 
\hline

LSTM[FastText] 
& 96.35 & 44.60 & 60.98 & 94.28 & 83.17 & 88.37 & 99.38 & 79.70 & 88.46 \\

LSTM [Glove] & 87.60 & 47.74 & 61.80 & 99.99 &83.85 & 91.21 & 95.95 & 82.18 & 88.53 \\

LSTM [Word2Vec] & 66.16 &54.13 & 59.55 & 96.50&84.26 & 89.96 & 98.10& 76.73& 86.11\\
LSTM [BERT]
& 90.01 & 90.31 & 90.16
& 99.27 & 97.80 & 98.53
& 100.0 & 82.67 & 90.51 \\

LSTM [QTyBERT] & 93.85 & 82.08 & 87.57 & 99.00 & 93.95 & 96.41 & 99.42 & 85.15 & 91.73 \\
\hline

GRU [FastText] 
& 66.12 & 55.01 & 60.05 & 98.94 & 83.93 & 90.81 & 98.20 & 81.19 & 88.89 \\
GRU  [Glove] & 81.34& 53.87& 64.82& 97.03& 85.06& 90.65 & 97.48 & 76.73 & 85.87\\
GRU [Word2Vec] & 60.62 & 66.43& 63.39 & 86.41 & 94.97& 90.49 & 98.73 & 77.23& 86.67\\
GRU [BERT]
& 89.16 & 92.73 & 90.91
& 99.39 & 93.97 & 96.60
& 100.0 & 83.67 & 91.11 \\

GRU [QTyBERT] & 94.16 & 85.04  & 89.36 & 99.43 & 93.29 & 96.26 & 98.30 & 85.64 & 91.53\\
\hline

CNN [FastText] 
& 61.61 & 58.63 & 60.08 & 99.96 & 94.26 & 97.02 & 97.08 & 82.17 & 89.00 \\

CNN [Glove] & 78.98 & 56.54 & 65.90 & 99.98& 94.51& 97.13 & 97.91 & 83.74& 90.27 \\

CNN [Word2Vec] & 81.65 & 53.59 & 64.71  & 96.26& 94.57& 95.41 & 100.0 & 81.68 & 89.91 \\
CNN [BERT]
& 96.06 & 66.19 & 78.37
& 99.52 & 93.62 & 96.48
& 100.0 & 83.66 & 91.10 \\

CNN [QTyBERT] & 98.16 & 69.42 & 79.52 & 99.71 & 93.46 & 96.49 & 97.18 & 85.15 & 90.76\\

\hline

RNN [FastText] 
& 47.95 & 64.64 & 55.05 & 96.82 & 79.36 & 87.22 & 98.75 & 78.21 & 87.29 \\

RNN [Glove] & 70.17& 50.48 & 58.72 & 95.95 & 82.17 & 88.53 & 94.15 & 79.70& 86.32\\

RNN [Word2Vec] & 48.52 & 66.49& 56.09 & 87.12& 94.20& 90.52 & 100.0 & 82.17 & 90.21\\
RNN [BERT]
& 86.21 & 56.63 & 68.36
& 96.67 & 93.63 & 95.12
& 100.0 & 82.67 & 90.51 \\

RNN [QTyBERT] & 93.65 & 86.42 & 
89.89 & 99.63 & 92.57 & 95.97 & 98.41 & 92.08 & 95.14\\
\hline

\end{tabular}
\end{table*}
To evaluate QTyBERT, we define the research question:
\begin{itemize}
    \item \textbf{RQ2.Performance}: How do QTyBERT perform compared to prior semantic log representation methods when serving input to downstream DL models?
    
\end{itemize}

To develop QTyBERT, we sample additional unlabeled log events from the same software systems in the USENIX CFDR repository used in our empirical study. Specifically, 
to build SysBE for each system $s_j$, we randomly select 70 unlabeled log events to form its calibration dataset $\mathcal{D}_{\text{cal}}^{(j)}$ to quantize the original TinyBERT $\mathcal{M}$ and obtain the corresponding SysBE $\mathcal{M}_{\text{q}}^{(j)}$; moreover, we randomly sample 25{,}000 unlabeled log events from each system, which together constitute the dataset $\mathcal{D}_{\text{cro}}$ to train CroSysEh. The sampled log events may overlap with the training set used in our empirical study, but are disjoint from the testing set used in our empirical study and occur earlier than the log events in this testing set to preserve chronological order. We construct SysBE on CPUs, and train CroSysEh on a GPU since running BERT on CPU is significantly slow for large-scale log data (see Table~\ref{tab:log_embedding_generation_time}), using the same hardware configuration as our empirical study (Section \ref{EmS_imple}).  As a result, each SysBE $\mathcal{M}_{\text{q}}^{(j)}$ has a storage footprint of 43 MB, which is substantially smaller than BERT ($\approx$440 MB), GloVe ($\approx$1 GB), and FastText ($\approx$4.51 GB), while remaining reasonably compact compared to system-specific Word2Vec models (1.68–10.44 MB). CroSysEh has a storage footprint of 968~KB.



We evaluate QTyBERT against the semantic log representation methods in our empirical study under the same experimental settings (Section \ref{sec:EmS_emperiment_setup}), i.e., using the same log datasets, DL models with fixed-size window strategy, training and testing sets, CPU deployment settings, implementation settings, 
and evaluation metrics. Specifically, for each system, we encode log events in training and testing sets into embeddings using the corresponding SysBE, and then map these embeddings to the final embedding space through CroSysEh. The final log embeddings are input to the DL models using the fixed-size window strategy for anomaly detection.

%% file: sections/results.tex
\section{QTyBERT Experiment Results and Analysis}

\subsection{RQ2. Performance} 

\subsubsection{Effectiveness} 
\label{sec:RQ2_log_representaion_effectviness}
Our QTyBERT generates effective log embeddings that are comparable to those of BERT, and even outperform it in certain cases. As shown in Table~\ref{tab:study_effet_results}, for most DL models, using log embedding from QTyBERT instead of BERT leads to F1-score differences within 1\%, either slightly higher or lower. A notable exception is RNN on BGL, where using QTyBERT yields a 21.53\% higher F1-score than BERT (89.89\% vs. 68.36\%), and achieves performance comparable to complex DL models TransEnc (89.98\%) and AttBiLSTM (91.77\%). Furthermore, with QTyBERT-based log embeddings, RNN achieves the highest F1-score on Spirit (95.14\%) outperforming all other DL models across different representation methods. These results indicate that QTyBERT generates effective log embeddings, enabling a vanilla RNN to achieve competitive detection effectiveness compared to more complex DL models.

\begin{table*}[!hb]
\setcounter{table}{3}
\centering
\caption{Log embedding generation time of different log representation methods. 
}
\begin{tabular}{llrrrrrr}
\toprule
\multirow{2}{*}{CPU} & \multirow{2}{*}{Method}
& \multicolumn{2}{c}{BGL} 
& \multicolumn{2}{c}{TB} 
& \multicolumn{2}{c}{Spirit} \\
\cmidrule(lr){3-4} \cmidrule(lr){5-6} \cmidrule(lr){7-8}
 & & Total (s) & Avg (ms) & Total (s) & Avg (ms) & Total (s) & Avg (ms) \\
\midrule
\multirow{5}{*}{8-core}
& FastText  & 111.88 & 0.24 & 130.67 & 0.09 & 110.23 & 0.22 \\
& GloVe     & 96.44  & 0.20 & 124.13 & 0.09 & 119.33 & 0.24 \\
& Word2Vec  & 54.98  & 0.12 & 67.50  & 0.05 & 57.87  & 0.12 \\
& BERT      & 2065.88 & 4.38 & 10392.57 & 7.44 & 3450.03 & 6.91 \\
& QTyBERT   & 167.66 & 0.36 & 504.22 & 0.36 & 178.20 & 0.36 \\
\midrule
\multirow{5}{*}{4-core}
& FastText  & 141.98 & 0.30 & 136.14 & 0.10 & 128.08 & 0.26 \\
& GloVe     & 98.13  & 0.21 & 126.74 & 0.09 & 120.76 & 0.24 \\
& Word2Vec  & 56.55  & 0.12 & 66.30  & 0.05 & 58.98  & 0.12 \\
& BERT      & 4210.95 & 8.93 & 21779.26 & 15.59 & 7613.55 & 15.25 \\
& QTyBERT   & 297.78 & 0.63 & 897.74 & 0.64 & 303.14 & 0.61 \\
\bottomrule
\end{tabular}
\label{tab:log_embedding_generation_time}
\end{table*}

\begin{figure}[!ht]
    \centering
    \begin{subfigure}{0.49\linewidth}
        \centering
        \includegraphics[width=\linewidth]{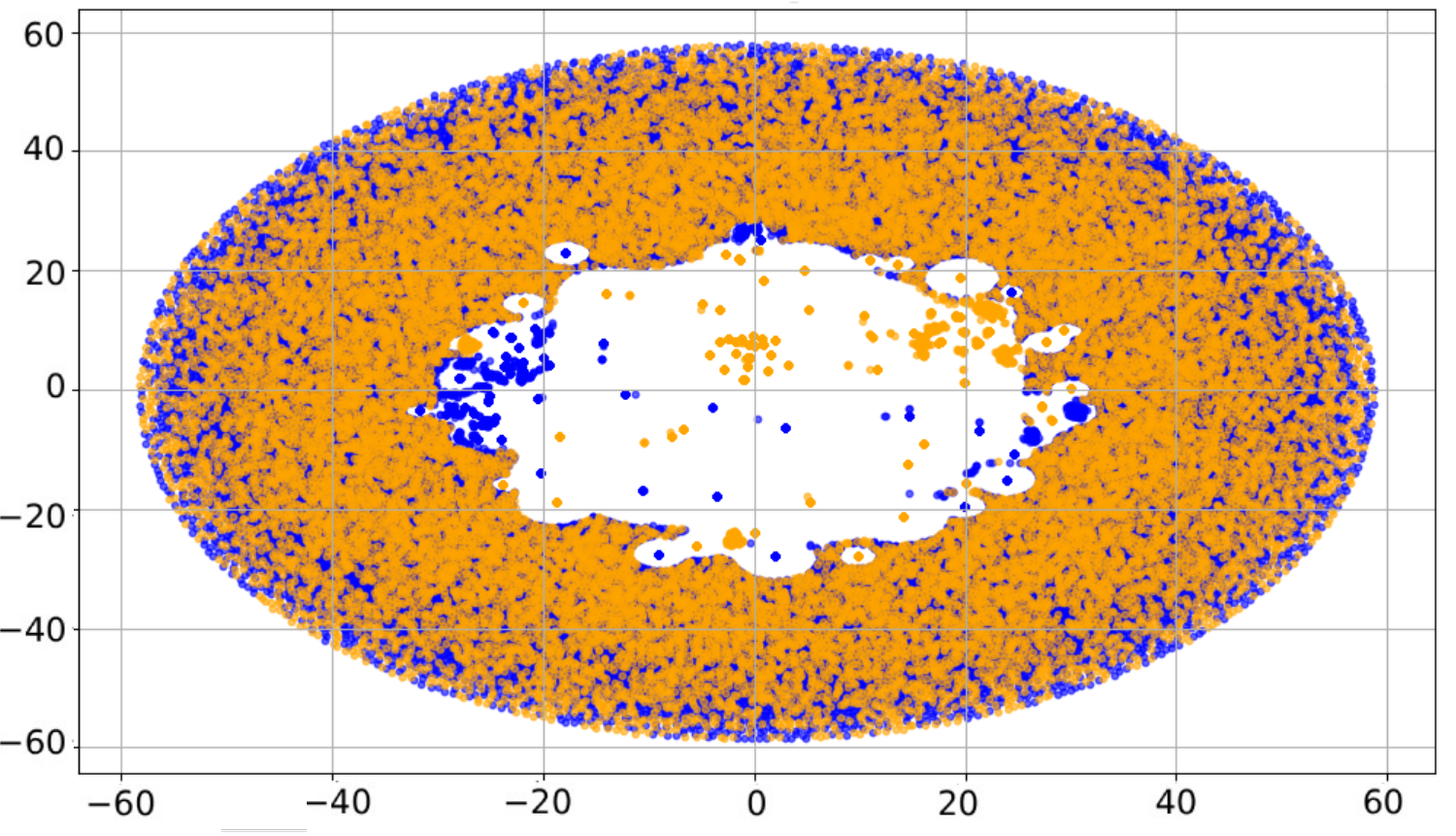}
        \caption{BGL}
    \end{subfigure}
    \hfill
    \begin{subfigure}{0.49\linewidth}
        \centering
        \includegraphics[width=\linewidth]{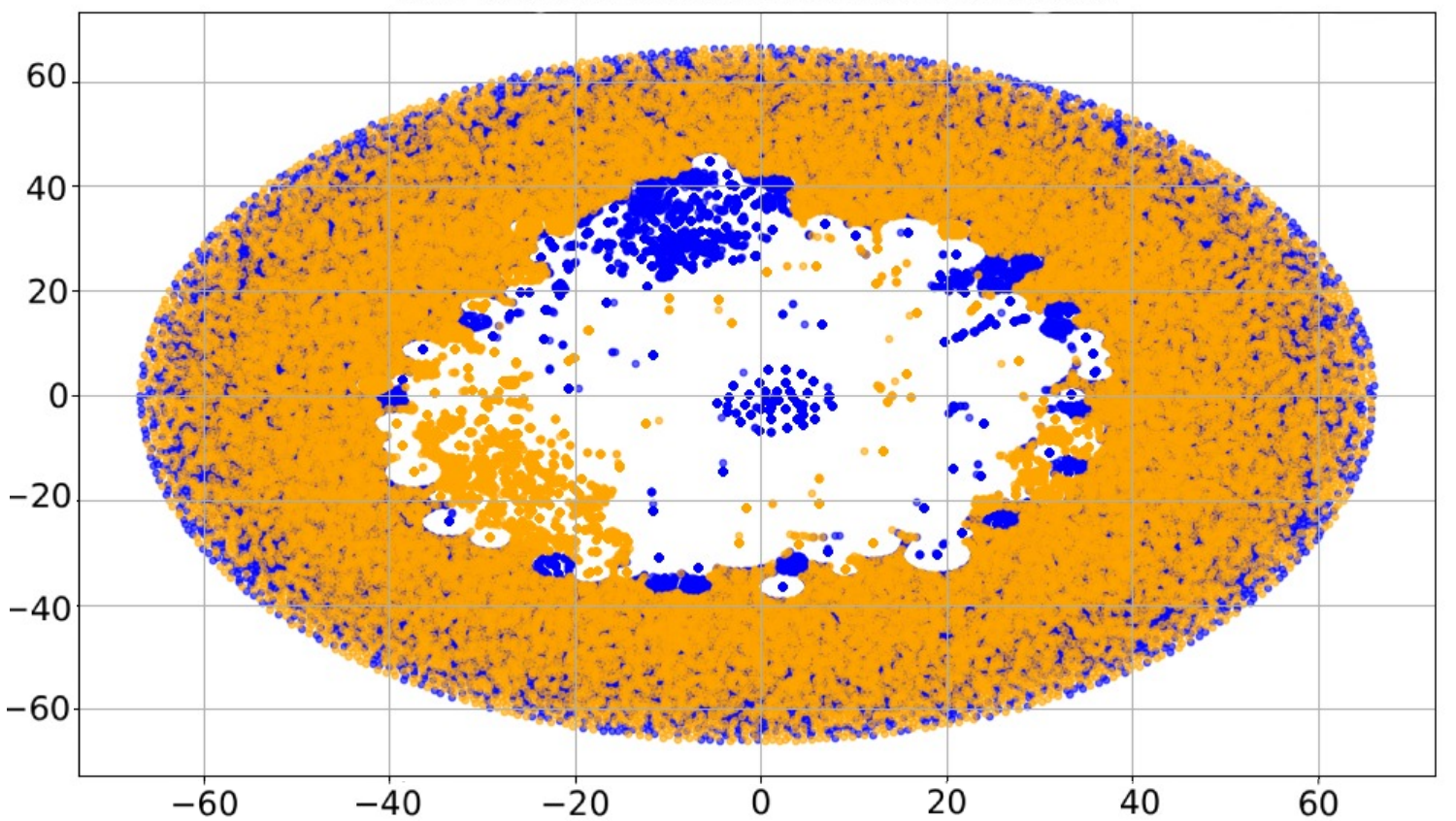}
        \caption{TB}
    \end{subfigure}
    
    \vspace{-0.3em}
    \begin{subfigure}{0.49\linewidth}
        \centering
        \includegraphics[width=\linewidth]{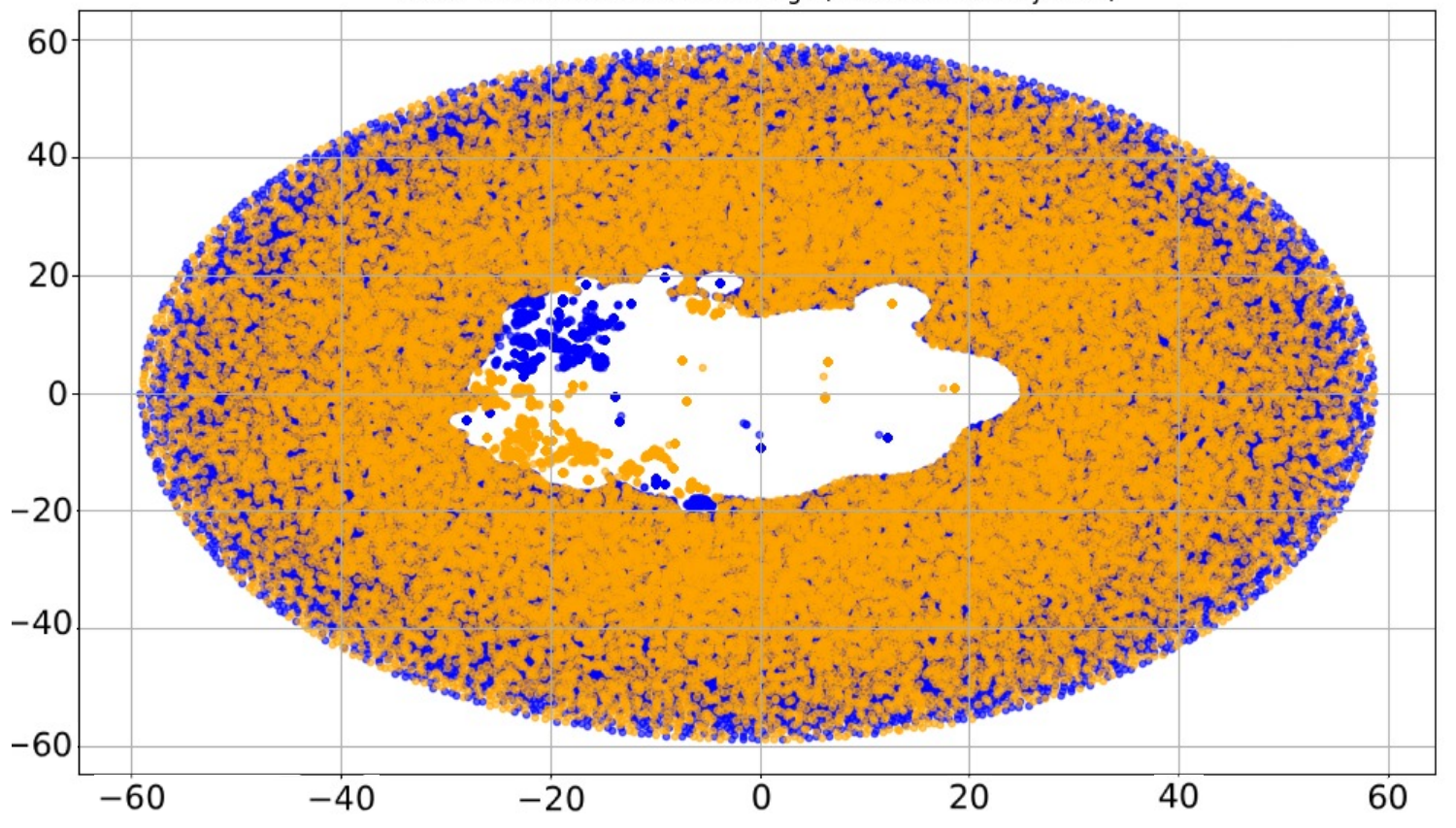}
        \caption{Spirit}
    \end{subfigure}
    \caption{t-SNE visualizations comparing log embeddings generated by {\color{bertblue}$\bullet$} BERT and 
{\color{qtyorange}$\bullet$} QTyBERT.}
    \label{fig:tsne_visualization}
    \Description{Visual}
\end{figure}
To investigate how QTyBERT learns from BERT through its CroSysEh, we perform both visualization (Figure~\ref{fig:tsne_visualization}) and quantitative comparison (Table~\ref{tab:cosine_spearman}) of their generated log embeddings. From each system, we randomly sample 50,000 log events and obtain their embeddings with BERT and QTyBERT, respectively. We then apply t-SNE \cite{vanDerMaaten2008} to project log embeddings of these two methods into a two-dimensional space. As shown in Figure~\ref{fig:tsne_visualization}, for each system, log embeddings generated by these two methods exhibit a high degree of overlap in structure, while maintaining some distributional differences. This is further supported by quantitative results in Table~\ref{tab:cosine_spearman}. Spearman correlation between log embeddings of QTyBERT and BERT is significantly high (0.6095--0.8089, $p < 0.001$), indicating that their log embeddings have similar structural relationships. The cosine similarity between their log embeddings is low (0.0492--0.1194). This is expected, as quantization in SysBE alters numerical values and CroSysEh learns the shared semantic structure across systems in the embedding space, which may result in different embedding direction and scale. These results indicate that QTyBERT learns the underlying functional semantic structure of BERT’s embedding space rather than replicating its exact embedding values.

\begin{table}[!ht]
\setcounter{table}{2}
\centering
\caption{Cosine Similarity and Spearman Correlation of Log Embeddings (QTyBERT vs. BERT)
}
\label{tab:cosine_spearman}
\begin{tabular}{lcc}
\toprule
System & Cosine (Mean) & Spearman $\rho$ \\
\midrule
BGL  & 0.0492 & 0.7383*** \\
TB & 0.1194 & 0.6095*** \\
SPIRIT & 0.1016 & 0.8089*** \\
\bottomrule
\end{tabular}
\begin{center}
\footnotesize ***p < 0.001; **p < 0.01; *p < 0.05
\end{center}
\vspace{-1em}
\end{table}

\begin{table*}[t]
\setcounter{table}{4}
\centering
\caption{
Detection latency (in seconds) of DL models using log embeddings from different semantic log representation methods
}
\label{tab:detection_latency}
\begin{tabular}{llrrrrrrr}
\toprule
\multirow{2}{*}{System} & \multirow{2}{*}{DL Model}
& \multicolumn{3}{c}{8-core} 
& \multicolumn{3}{c}{4-core} \\
\cmidrule(lr){3-5} \cmidrule(lr){6-8}
 & & Static Avg ($\Delta_{\max}$) & BERT & QTyBERT & Static Avg ($\Delta_{\max}$) & BERT & QTyBERT \\
\hline
\multirow{6}{*}{BGL}
& TransEnc        & 22.63 (1.20) & 25.61 & 24.06 & 47.77 (3.77) & 54.21 & 51.42 \\
& BiLSTM+WgtAttn  & 3.94 (1.18)  & 4.66  & 4.40  & 5.61 (0.94)  & 6.99  & 5.87  \\
& LSTM            & 2.01 (0.43)  & 2.23  & 2.32  & 3.05 (0.40)  & 3.25  & 3.75  \\
& GRU             & 2.65 (1.13)  & 4.01  & 3.35  & 3.52 (0.98)  & 4.85  & 4.44  \\
& CNN             & 1.37 (0.42)  & 1.65  & 1.39  & 1.47 (0.42)  & 1.77  & 1.47  \\
& RNN             & 0.74 (0.15)  & 0.80  & 0.79  & 0.77 (0.04)  & 0.82  & 0.81  \\
\midrule
\multirow{6}{*}{TB}
& TransEnc        & 56.63 (2.76) & 72.13 & 72.56 & 132.39 (7.40) & 161.32 & 160.30 \\
& BiLSTM+WgtAttn  & 4.22 (0.47)  & 7.99  & 7.87  & 8.04 (0.14)   & 15.10  & 14.03  \\
& LSTM            & 5.40 (0.11)  & 5.76  & 5.36  & 7.08 (0.19)   & 7.72   & 7.25   \\
& GRU             & 4.02 (0.14)  & 4.51  & 4.63  & 7.45 (0.54)   & 10.25  & 9.33   \\
& CNN             & 2.69 (0.27)  & 2.97  & 2.77  & 2.81 (0.30)   & 3.00   & 2.80   \\
& RNN             & 2.51 (0.05)  & 2.70  & 2.70  & 2.60 (0.12)   & 2.84   & 2.78   \\
\midrule
\multirow{6}{*}{Spirit}
& TransEnc        & 20.89 (3.21) & 25.00 & 25.21 & 45.69 (3.43) & 55.13 & 55.86 \\
& BiLSTM+WgtAttn  & 4.12 (0.43)  & 5.86  & 5.86  & 6.02 (0.13)  & 7.26  & 7.96  \\
& LSTM            & 1.95 (0.17)  & 2.13  & 2.11  & 2.95 (0.23)  & 3.10  & 3.00  \\
& GRU             & 2.86 (0.18)  & 3.27  & 3.30  & 3.86 (0.37)  & 4.11  & 4.17  \\
& 1D-CNN          & 1.28 (0.08)  & 1.50  & 1.53  & 1.95 (0.10)  & 2.10  & 2.08  \\
& RNN             & 1.01 (0.01)  & 1.09  & 1.05  & 1.04 (0.08)  & 1.08  & 1.06  \\
\bottomrule
\end{tabular}
\end{table*}


\subsubsection{Efficiency}

\begin{figure}[!htp]
\begin{tikzpicture}

\begin{axis}[
    tradeoff style,
    name=bgl,
    title={BGL},
    xmode=log,
    ymin=56, ymax=93,
    ytick={60, 65, 70, 75, 80, 85, 90},
]
\addplot[only marks, mark=*, mark size=2.5pt, color=staticcol!70!black] coordinates {(0.1862, 62.30)};
\addplot[only marks, mark=diamond*, mark size=3pt, color=staticcol, opacity=0.4] coordinates {(0.2098, 62.30)};
\addplot[only marks, mark=*, mark size=2.5pt, color=bertcol!70!black] coordinates {(4.3829, 84.14)};
\addplot[only marks, mark=diamond*, mark size=3pt, color=bertcol, opacity=0.4] coordinates {(8.9338, 84.14)};
\addplot[only marks, mark=*, mark size=2.5pt, color=qtycol!70!black] coordinates {(0.3557, 88.02)};
\addplot[only marks, mark=diamond*, mark size=3pt, color=qtycol, opacity=0.4] coordinates {(0.6318, 88.02)};
\node[font=\tiny, above=2pt, color=staticcol] at (axis cs:0.1862, 62.30) {Static};
\node[font=\tiny, above=2pt, color=bertcol] at (axis cs:4.3829, 84.14) {BERT};
\node[font=\tiny, above=2pt, color=qtycol] at (axis cs:0.3557, 88.02) {QTyBERT};
\end{axis}

\begin{axis}[
    tradeoff style,
    name=tb,
    anchor=west,
    at={($(bgl.east)+(1cm,0)$)},
    title={TB},
    xmode=log,
    ymin=90, ymax=98,
    ylabel={},
]
\addplot[only marks, mark=*, mark size=2.5pt, color=staticcol!70!black] coordinates {(0.0769, 92.67)};
\addplot[only marks, mark=diamond*, mark size=3pt, color=staticcol, opacity=0.4] coordinates {(0.0786, 92.67)};
\addplot[only marks, mark=*, mark size=2.5pt, color=bertcol!70!black] coordinates {(7.4406, 95.89)};
\addplot[only marks, mark=diamond*, mark size=3pt, color=bertcol, opacity=0.4] coordinates {(15.5928, 95.89)};
\addplot[only marks, mark=*, mark size=2.5pt, color=qtycol!70!black] coordinates {(0.3610, 95.56)};
\addplot[only marks, mark=diamond*, mark size=3pt, color=qtycol, opacity=0.4] coordinates {(0.6427, 95.56)};
\node[font=\tiny, below=2pt, color=staticcol] at (axis cs:0.0769, 92.67) {Static};
\node[font=\tiny, above=2pt, color=bertcol] at (axis cs:7.4406, 95.89) {BERT};
\node[font=\tiny, above=2pt, color=qtycol] at (axis cs:0.3610, 95.56) {QTyBERT};
\end{axis}

\begin{axis}[
    tradeoff style,
    name=spirit,
    anchor=north,
    at={($(bgl.south)!0.5!(tb.south)+(0,-1.2cm)$)},
    title={Spirit},
    xmode=log,
    ymin=84, ymax=94,
    ytick={85, 87, 89, 91, 93},
]
\addplot[only marks, mark=*, mark size=2.5pt, color=staticcol!70!black] coordinates {(0.1920, 87.40)};
\addplot[only marks, mark=diamond*, mark size=3pt, color=staticcol, opacity=0.4] coordinates {(0.2056, 87.40)};
\addplot[only marks, mark=*, mark size=2.5pt, color=bertcol!70!black] coordinates {(6.9126, 90.39)};
\addplot[only marks, mark=diamond*, mark size=3pt, color=bertcol, opacity=0.4] coordinates {(15.2547, 90.39)};
\addplot[only marks, mark=*, mark size=2.5pt, color=qtycol!70!black] coordinates {(0.3570, 91.68)};
\addplot[only marks, mark=diamond*, mark size=3pt, color=qtycol, opacity=0.4] coordinates {(0.6074, 91.68)};
\node[font=\tiny, below=2pt, color=staticcol] at (axis cs:0.1920, 87.40) {Static};
\node[font=\tiny, above=2pt, color=bertcol] at (axis cs:6.9126, 90.39) {BERT};
\node[font=\tiny, above=2pt, color=qtycol] at (axis cs:0.3570, 91.68) {QTyBERT};
\end{axis}


\end{tikzpicture}
\vspace{-0.5cm}
\caption{Trade-off between detection effectiveness (Avg F1-score \%) and log embedding generation efficiency (ms/Log). \textcolor{staticcol}{$\bullet$}\textcolor{staticcol}{$\diamond$}~Static, \textcolor{qtycol}{$\bullet$}\textcolor{qtycol}{$\diamond$}~QTyBERT, \textcolor{bertcol}{$\bullet$}\textcolor{bertcol}{$\diamond$}~BERT; $\bullet$~8-core CPU, $\diamond$~4-core CPU.}
\label{fig:tradeoff}
\Description{Trade off}
\vspace{-1.3em}
\end{figure}
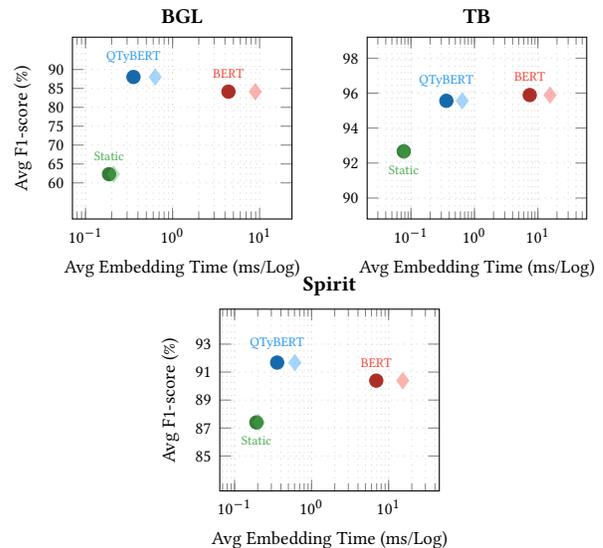

\textbf{Log Embedding Generation.} QTyBERT generates log embedding significantly faster than BERT in CPU-only deployment settings across all three systems, as shown in Table~\ref{tab:log_embedding_generation_time}. On the 8-core CPU setting, QTyBERT is 12$\times$ to 21$\times$ faster than BERT, with an average generation time of 0.36 ms per log event compared to 4.38--7.44 ms per log event for BERT. On the 4-core CPU setting, the speedup is even more pronounced, with QTyBERT being 14$\times$ to 25$\times$ faster than BERT, achieving 0.61--0.64 ms per log event compared to 8.93--15.59 ms per log event for BERT. These correspond to more than a 94\% reduction in embedding generation time on both CPU settings. For example, on TB with over 1.39 million log events, the total log embedding generation time is reduced from more than 10{,}300 seconds ($\approx$2.9 hours) with BERT to about 500 seconds ($\approx$8 minutes) with QTyBERT on 8 CPU cores, and from over 21{,}700 seconds ($\approx$6 hours) with BERT to under 900 seconds ($\approx$15 minutes) with QTyBERT on 4 CPU cores. 
Compared to static embedding methods (FastText, GloVe, and Word2Vec), QTyBERT is still slower, but also achieves sub-millisecond latency per log event across all systems and CPU configurations.

\textbf{Detection latency.} As shown in Table \ref{tab:detection_latency}, the detection latency is highly consistent when using log embeddings from QTyBERT and BERT, with differences of less than 5\% in most cases across DL models and systems. This is expected, as QTyBERT preserves the same embedding dimensionality as BERT (Section \ref{sec:emb_distillation}), leading to similar processing times for downstream DL models.

\subsubsection{Trade-Off} Figure 3 plots, for each system and each representation method, the average F1-score across DL models against the average embedding generation time per log event under both CPU settings. Here, “Static” denotes the average results of static word embedding methods (Word2Vec, GloVe, and FastText). As Figure~\ref{fig:tradeoff} shows, 
QTyBERT achieves a better trade-off between detection effectiveness and log embedding generation efficiency compared to static word embedding and BERT methods.

\subsection{Training Costs}

As shown in Table~\ref{tab:training_costs}, the training cost of QTyBERT consists of two components. First, obtaining SysBE for each target system requires only about 0.05 seconds under both CPU settings. Second, CroSysEh is trained once for all systems. As CroSysEh is lightweight, optimizing its parameters takes only about 7 seconds. The overall training cost of CroSysEh is approximately 289 seconds ($\approx$ 4.8 minutes), dominated by log embedding generation for $\mathcal{D}_{\text{cro}}$ using BERT and TinyBERT. Importantly, this cost is incurred only once. During deployment, QTyBERT reduces embedding generation time by approximately 94\% compared to BERT, while maintaining comparable anomaly detection effectiveness for downstream DL models. In production environments where logs are continuously generated at a large scale, the resulting recurring savings in embedding generation will quickly outweigh this one-time training cost.

\begin{table}[!ht]
\centering
\caption{Training cost of QTyBERT.}
\label{tab:training_costs}
\begin{tabular}{llr}
\hline
Component & Setting & Time (s) \\
\hline
SysBE & BGL, 8/4-core CPU & 0.05 / 0.47 \\
 & TB, 8/4-core CPU & 0.05 / 0.47 \\
 & Spirit, 8/4-core CPU & 0.05 / 0.47 \\
\midrule
 BERT & GPU & 218.41 \\
TinyBERT & GPU & 63.85 \\ 
CroSysEh & 80 epochs, GPU & $\sim$7.16 (0.09/epoch) \\
\hline
\end{tabular}
\vspace{-1em}
\end{table}

\subsection{Ablation study} 
\label{sec:ablation_study}

We conduct an ablation study using RNN as the downstream DL model, as it has the lowest detection latency (Table~\ref{tab:detection_latency}) and exhibits the highest effectiveness gains with QTyBERT-based log embeddings among all DL models (Table~\ref{tab:study_effet_results}). 
Table ~\ref{tab:ablation_effectiveness} and~Table \ref{tab:ablation_efficiency} report the ablation results on detection effectiveness and log embedding generation efficiency, respectively.

\textbf{CroSysEh.} Removing CroSysEh (w/o CroSysEh) leads to F1-score drops on all systems: BGL (-9.73\%), TB (-2.89\%), and Spirit (-4.72\%). This confirms that CroSysEh improves the effectiveness of log embeddings generated by SysBE for anomaly detection. Meanwhile, CroSysEh adds less than 0.6\%  to the total embedding generation time across all systems and CPU settings, meaning the effectiveness gains come with only a marginal increase in computational cost. Replacing cross-system training with a single-system variant (w/ sig.CroSysEh) in CroSysEh yields slightly higher F1-scores (0.29-2.1\%) on all systems, suggesting that single-system training can better fit system-specific patterns. Cross-system training learns from logs of multiple systems, trading a small amount of dataset-specific performance for a shared CroSysEh reusable across systems without retraining. This is more practical for organizations operating multiple systems.

\textbf{SysBE.} Removing SysBE (w/o SysBE) causes only minor changes in F1-scores but significantly increases embedding generation time by around 3\%-20\% across all systems, indicating that SysBE’s quantization substantially improves efficiency while having little impact on downstream anomaly detection effectiveness. However, removing the calibration step (w/o calibration) in SysBE causes dramatic drops in F1-score across all systems: BGL ($-$19.19\%), TB ($-$4.68\%), and Spirit ($-$14.88\%).  This confirms that system-specific calibration is essential during quantization for preserving the embedding quality. As shown in Table~\ref{tab:calibration_samples}, F1-scores drop notably when fewer than 70 calibration samples are used, indicating that insufficient calibration samples fail to adequately cover the target system’s activation distribution, which in turn degrades quantization quality and detection effectviness.

\begin{table}[h]
\caption{Ablation study on effectiveness (F1-score, \%).}
\begin{tabular}{p{2.22cm}p{1.32cm}lp{1.32cm}}
\toprule
Method & BGL \footnotesize{($\Delta$)} & TB \footnotesize{($\Delta$)} & Spirit \footnotesize{($\Delta$)} \\
\midrule
QTyBERT & 89.89 & 95.97 & 95.14 \\
w/o CroSysEh & 80.16 \scriptsize{(-9.73)} & 93.08 \scriptsize{(-2.89)} & 91.20 \scriptsize{(-4.72)} \\
w/ sig.CroSysEh & 90.18 \scriptsize{(+0.29)} & 96.24 \scriptsize{(+0.27)} & 97.24 \scriptsize{(+2.10)} \\
w/o SysBE & 90.59 \scriptsize{(+0.70)} & 96.09 \scriptsize{(+0.12)} & 95.41 \scriptsize{(+0.27)} \\
w/o calibration & 70.70 \scriptsize{(-19.19)} & 91.29 \scriptsize{(-4.68)} & 80.26 \scriptsize{(-14.88)} \\
\bottomrule
\end{tabular}
\label{tab:ablation_effectiveness}
\end{table}

\begin{table}[!ht]
\caption{Ablation study on log embedding generation efficiency (in seconds).}
\label{tab:ablation_efficiency}
\begin{tabular}{p{0.68cm}p{0.85cm}p{1.25cm}p{2.3cm}p{1.67cm}}
\toprule
 & & \multicolumn{3}{c}{Log Embedding Generation Time (s)} \\
\cmidrule(lr){3-5}
 & CPU & QTyBERT & w/o CroSysEh \scriptsize{($\Delta$)} & w/o SysBE \scriptsize{($\Delta$) }\\
\midrule
\multirow{2}{*}{BGL}
& 8-core & 167.66 & 166.63 \scriptsize{(-1.03)} & 187.79 \scriptsize{(+20.13)} \\
& 4-core & 297.78 & 296.01 \scriptsize{(-1.77)} & 356.56 \scriptsize{(+58.78)} \\
\midrule
\multirow{2}{*}{TB}
& 8-core & 504.22 & 501.24 \scriptsize{(-2.98)} & 557.68 \scriptsize{(+53.46)} \\
& 4-core & 897.74 & 892.76 \scriptsize{(-4.98)} & 928.21 \scriptsize{(+30.47)} \\
\midrule
\multirow{2}{*}{Spirit}
& 8-core & 178.20 & 177.18 \scriptsize{(-1.02)} & 186.78 \scriptsize{(+8.58)} \\
& 4-core & 303.14 & 301.29 \scriptsize{(-1.85)} & 320.98 \scriptsize{(+17.84)} \\
\bottomrule
\end{tabular}
\end{table}

\begin{table}[ht]
\centering
\caption{Effect of calibration sample size (F1-score, \%) 
} 
\label{tab:calibration_samples}
\begin{tabular}{lccc}
\toprule
N of log events & BGL & TB & Spirit \\
\midrule
30  & 72.41 & 91.84 & 72.89 \\
50 & 70.69 & 90.26 & 87.29\\
70 (ours)  & 89.89 & 95.97 & 95.14 \\
100 & 89.99 & 94.98 & 93.50 \\
\bottomrule
\end{tabular}
\end{table}

%% file: sections/conclution.tex
\section{Threats to Validity}
A threat to construct validity is that some DL models were originally designed for session-level anomaly detection. By studying prior settings \cite{CroSysLog,NeuralLog2021,Adha2025}, we find that both session-level and event-level detection operate on windowed log sequences and differ only in prediction granularity. Therefore, applying these DL models to our setting primarily requires adapting the prediction target. 

One threat to internal validity concerns the construction of the calibration dataset. In our experiments, we randomly sample 70 unlabeled log events from each target system, which yields high effectiveness across all three systems. However, prior work has shown that random calibration data selection may introduce performance instability due to activation distribution mismatch \cite{zhang2020qbert}, and more principled selection strategies may further improve calibration quality. We mitigate this threat by using system-specific log events for calibration, ensuring the calibration data reflects the actual activation distribution of the target system. 

A potential threat to external validity lies in our evaluation. Our experiments were conducted on publicly available datasets of large-scale supercomputing systems. While these real-world datasets are widely used in prior work to ensure fair comparison, production environments of different software systems may introduce additional diversity and complexity, due to the heterogeneous nature of software systems and varied logging practices. Expanding the evaluation to more software systems and incorporating feedback from practitioners would provide complementary insights.

\section{Conclusion}
This paper makes contributions to semantic log representation for DL-based log event-level anomaly detection. First, we conduct a comprehensive empirical study benchmarking widely used semantic log representation methods across a broad set of DL models under CPU-only deployment settings using publicly available log datasets. We identify a clear trade-off between static word embedding methods and the BERT-based contextual embedding method in detection effectiveness and log embedding generation efficiency. Second, motivated by this finding, we propose QTyBERT, a novel semantic log representation method that better balances this trade-off. Future work will aim to improve the generalizability and interpretability of
QTyBERT. We are seeking opportunities to extend its evaluation using log datasets from our
local supercomputing center, which will allow us to study its performance under more diverse
operational conditions. We also plan to collaborate with practitioners to assess its practical usage in real-world practices. Their feedback will guide subsequent enhancements to improve the usability.

\section{Data Availability}
The datasets used in this paper are publicly available and can be accessed from their original sources as cited in the paper. Upon acceptance, we will make this package publicly available.

\section{Acknowledgment}
This work is funded by the EuroHPC Joint Undertaking and its members, including top-up funding by the Ministry of Education and Culture. The work is supported by the Research
Council of Finland (grant id: 359861, the MuFAno project). The authors acknowledge CSC-IT Center for Science, Finland, for providing computational resources.

%% file: mainbib.bib
@article{hashemi2024onelog,
  title={Onelog: towards end-to-end software log anomaly detection},
  author={Hashemi, Shayan and M{\"a}ntyl{\"a}, Mika},
  journal={Automated Software Engineering},
  volume={31},
  number={2},
  pages={37},
  year={2024},
  publisher={Springer}
}

@article{Bertcompress,
    title = "Compressing Large-Scale Transformer-Based Models: A Case Study on {BERT}",
    author = "Ganesh, Prakhar  and
      Chen, Yao  and
      Lou, Xin  and
      Khan, Mohammad Ali  and
      Yang, Yin  and
      Sajjad, Hassan  and
      Nakov, Preslav  and
      Chen, Deming  and
      Winslett, Marianne",
    editor = "Roark, Brian  and
      Nenkova, Ani",
    journal = "Transactions of the Association for Computational Linguistics",
    volume = "9",
    year = "2021",
    address = "Cambridge, MA",
    publisher = "MIT Press",
    url = "https://aclanthology.org/2021.tacl-1.63/",
    doi = "10.1162/tacl_a_00413",
    pages = "1061--1080"
}

@ARTICLE{SwissLog,
  author={Li, Xiaoyun and Chen, Pengfei and Jing, Linxiao and He, Zilong and Yu, Guangba},
  journal={IEEE Transactions on Dependable and Secure Computing}, 
  title={SwissLog: Robust Anomaly Detection and Localization for Interleaved Unstructured Logs}, 
  year={2023},
  volume={20},
  number={4},
  pages={2762-2780},
  doi={10.1109/TDSC.2022.3162857}}

@ARTICLE{GRUBaseline,
  author={Studiawan, Hudan and Sohel, Ferdous and Payne, Christian},
  journal={IEEE Transactions on Dependable and Secure Computing}, 
  title={Anomaly Detection in Operating System Logs with Deep Learning-Based Sentiment Analysis}, 
  year={2021},
  volume={18},
  number={5},
  pages={2136-2148},
  doi={10.1109/TDSC.2020.3037903}}

@article{Cerqueira2020,
  author    = {Vitor Cerqueira and Lu{\'i}s Torgo and Igor Mozeti{\v{c}}},
  title     = {Evaluating time series forecasting models: an empirical study on performance estimation methods},
  journal   = {Machine Learning},
  volume    = {109},
  number    = {11},
  pages     = {1997--2028},
  year      = {2020},
  month     = nov,
  doi       = {10.1007/s10994-020-05910-7},
  url       = {https://doi.org/10.1007/s10994-020-05910-7}
}

@misc{MiniLM,
      title={MiniLM: Deep Self-Attention Distillation for Task-Agnostic Compression of Pre-Trained Transformers}, 
      author={Wenhui Wang and Furu Wei and Li Dong and Hangbo Bao and Nan Yang and Ming Zhou},
      year={2020},
      eprint={2002.10957},
      archivePrefix={arXiv},
      primaryClass={cs.CL},
      url={https://arxiv.org/abs/2002.10957}, 
}

@article{wu2016google,
  title={Google's neural machine translation system: Bridging the gap between human and machine translation},
  author={Wu, Yonghui and Schuster, Mike and Chen, Zhifeng and Le, Quoc V and Norouzi, Mohammad and Macherey, Wolfgang and Krikun, Maxim and Cao, Yuan and Gao, Qin and Macherey, Klaus and others},
  journal={arXiv preprint arXiv:1609.08144},
  year={2016},
  url={https://arxiv.org/abs/1609.08144}
}

@inproceedings{glove,
  title={GloVe: Global Vectors for Word Representation},
  author={Pennington, Jeffrey and Socher, Richard and Manning, Christopher D.},
  booktitle={Proceedings of the 2014 Conference on Empirical Methods in Natural Language Processing (EMNLP)},
  pages={1532--1543},
  year={2014},
  organization={Association for Computational Linguistics},
  doi={10.3115/v1/D14-1162}
}

@article{ParsingError2023,
    journal = {Empirical Software Engineering},
    author = {Fu, Ying and Yan, Meng and Xu, Zhou and Xia, Xin and Zhang, Xiaohong and Yang, Dan},
    title = {An empirical study of the impact of log parsers on the performance of log-based anomaly detection},
    year = {2022},
    issue_date = {Jan 2023},
    publisher = {Kluwer Academic Publishers},
    address = {USA},
    volume = {28},
    number = {1},
    issn = {1382-3256},
    url = {https://doi.org/10.1007/s10664-022-10214-6},
    doi = {10.1007/s10664-022-10214-6},
    month = nov,
    numpages = {39}
}

@inproceedings{LogAnomaly,
    author = {Meng, Weibin and Liu, Ying and Zhu, Yichen and Zhang, Shenglin and Pei, Dan and Liu, Yuqing and Chen, Yihao and Zhang, Ruizhi and Tao, Shimin and Sun, Pei and Zhou, Rong},
    title = {Loganomaly: unsupervised detection of sequential and quantitative anomalies in unstructured logs},
    year = {2019},
    isbn = {9780999241141},
    publisher = {AAAI Press},
    booktitle = {Proceedings of the 28th International Joint Conference on Artificial Intelligence},
    pages = {4739–4745},
    numpages = {7},
    location = {Macao, China},
    series = {IJCAI'19}
}

@article{LEE2023,
    title = {LAnoBERT: System log anomaly detection based on BERT masked language model},
    journal = {Applied Soft Computing},
    volume = {146},
    pages = {110689},
    year = {2023},
    issn = {1568-4946},
    doi = {https://doi.org/10.1016/j.asoc.2023.110689},
    url = {https://www.sciencedirect.com/science/article/pii/S156849462300707X},
    author = {Yukyung Lee and Jina Kim and Pilsung Kang}
}

@article{vanDerMaaten2008,
  author  = {Laurens van der Maaten and Geoffrey Hinton},
  title   = {Visualizing Data using t-SNE},
  journal = {Journal of Machine Learning Research},
  volume  = {9},
  number  = {Nov},
  pages   = {2579--2605},
  year    = {2008},
  url     = {http://www.jmlr.org/papers/v9/vandermaaten08a.html}
}

@article{HitAnomaly,
author = {Huang, Shaohan and Liu, Yi and Fung, Carol and He, Rong and Zhao, Yining and Yang, Hailong and Luan, Zhongzhi},
title = {HitAnomaly: Hierarchical Transformers for Anomaly Detection in System Log},
year = {2020},
issue_date = {Dec. 2020},
publisher = {IEEE Press},
volume = {17},
number = {4},
issn = {1932-4537},
url = {https://doi.org/10.1109/TNSM.2020.3034647},
doi = {10.1109/TNSM.2020.3034647},
journal = {IEEE Trans. on Netw. and Serv. Manag.},
month = dec,
pages = {2064–2076},
numpages = {13}
}

@article{LiHeng2023,
    author = {Wu, Xingfang and Li, Heng and Khomh, Foutse},
    title = {On the effectiveness of log representation for log-based anomaly detection},
    year = {2023},
    issue_date = {Nov 2023},
    publisher = {Kluwer Academic Publishers},
    address = {USA},
    volume = {28},
    number = {6},
    issn = {1382-3256},
    url = {https://doi.org/10.1007/s10664-023-10364-1},
    doi = {10.1007/s10664-023-10364-1},
    journal = {Empirical Softw. Engg.},
    month = oct,
    numpages = {39}
}

@article{Adha2025,
    author = {Hrusto, Adha and Ali, Nauman Bin and Engstr\"{o}m, Emelie and Wang, Yuqing},
    title = {Monitoring data for Anomaly Detection in Cloud-Based Systems: A Systematic Mapping Study},
    year = {2025},
    publisher = {Association for Computing Machinery},
    address = {New York, NY, USA},
    issn = {1049-331X},
    url = {https://doi.org/10.1145/3744556},
    doi = {10.1145/3744556},
    note = {Just Accepted},
    journal={ACM Transactions on Software Engineering and Methodology},
    month = jun
}

@INPROCEEDINGS{Oliner2007,
  author={Oliner, Adam and Stearley, Jon},
  booktitle={37th Annual IEEE/IFIP International Conference on Dependable Systems and Networks (DSN'07)}, 
  title={What Supercomputers Say: A Study of Five System Logs}, 
  year={2007},
  volume={},
  number={},
  pages={575-584},
  doi={10.1109/DSN.2007.103}}

@INPROCEEDINGS{LogLead,
  author={Mäntylä, Mika V. and Wang, Yuqing and Nyyssölä, Jesse},
  booktitle={2024 IEEE International Conference on Software Analysis, Evolution and Reengineering (SANER)}, 
  title={LogLead - Fast and Integrated Log Loader, Enhancer, and Anomaly Detector}, 
  year={2024},
  volume={},
  number={},
  pages={395-399},
  doi={10.1109/SANER60148.2024.00046}}

@misc{tinybert,
      title={TinyBERT: Distilling BERT for Natural Language Understanding}, 
      author={Xiaoqi Jiao and Yichun Yin and Lifeng Shang and Xin Jiang and Xiao Chen and Linlin Li and Fang Wang and Qun Liu},
      year={2020},
      eprint={1909.10351},
      archivePrefix={arXiv},
      primaryClass={cs.CL},
      url={https://arxiv.org/abs/1909.10351}, 
}

@inproceedings{le2022log,
author = {Le, Van-Hoang and Zhang, Hongyu},
title = {Log-based anomaly detection with deep learning: how far are we?},
year = {2022},
isbn = {9781450392211},
publisher = {Association for Computing Machinery},
address = {New York, NY, USA},
url = {https://doi.org/10.1145/3510003.3510155},
doi = {10.1145/3510003.3510155},
booktitle = {Proceedings of the 44th International Conference on Software Engineering},
pages = {1356–1367},
numpages = {12},
location = {Pittsburgh, Pennsylvania},
series = {ICSE '22}
}

@inproceedings{NeuralLog2021,
  title={Log-based anomaly detection without log parsing},
  author={Le, Van-Hoang and Zhang, Hongyu},
  booktitle={2021 36th IEEE/ACM International Conference on Automated Software Engineering (ASE)},
  pages={492--504},
  year={2021},
  publisher={IEEE},
  doi={10.1109/ASE51524.2021.9678773}
}

@Article{Baseline1dCNN,
AUTHOR = {Qazi, Emad Ul Haq and Almorjan, Abdulrazaq and Zia, Tanveer},
TITLE = {A One-Dimensional Convolutional Neural Network (1D-CNN) Based Deep Learning System for Network Intrusion Detection},
JOURNAL = {Applied Sciences},
VOLUME = {12},
YEAR = {2022},
NUMBER = {16},
ARTICLE-NUMBER = {7986},
URL = {https://www.mdpi.com/2076-3417/12/16/7986},
ISSN = {2076-3417},
DOI = {10.3390/app12167986}
}

@inproceedings{LogRobust,
author = {Zhang, Xu and Xu, Yong and Lin, Qingwei and Qiao, Bo and Zhang, Hongyu and Dang, Yingnong and Xie, Chunyu and Yang, Xinsheng and Cheng, Qian and Li, Ze and Chen, Junjie and He, Xiaoting and Yao, Randolph and Lou, Jian-Guang and Chintalapati, Murali and Shen, Furao and Zhang, Dongmei},
title = {Robust log-based anomaly detection on unstable log data},
year = {2019},
isbn = {9781450355728},
publisher = {Association for Computing Machinery},
address = {New York, NY, USA},
url = {https://doi.org/10.1145/3338906.3338931},
doi = {10.1145/3338906.3338931},
booktitle = {Proceedings of the 2019 27th ACM Joint Meeting on European Software Engineering Conference and Symposium on the Foundations of Software Engineering},
pages = {807–817},
numpages = {11},
location = {Tallinn, Estonia},
series = {ESEC/FSE 2019}
}

@INPROCEEDINGS{CroSysLog,
  author={Wang, Yuqing and Mäntylä, Mika V. and Nyyssölä, Jesse and Ping, Ke and Wang, Liqiang},
  booktitle={2025 IEEE International Conference on Software Analysis, Evolution and Reengineering (SANER)}, 
  title={Cross-System Software Log-based Anomaly Detection Using Meta-Learning}, 
  year={2025},
  volume={},
  number={},
  pages={454-464},
  doi={10.1109/SANER64311.2025.00049}}

@INPROCEEDINGS{LiqiangCNN,
  author={Lu, Siyang and Wei, Xiang and Li, Yandong and Wang, Liqiang},
  booktitle={2018 IEEE 16th Intl Conf on Dependable, Autonomic and Secure Computing, 16th Intl Conf on Pervasive Intelligence and Computing, 4th Intl Conf on Big Data Intelligence and Computing and Cyber Science and Technology Congress(DASC/PiCom/DataCom/CyberSciTech)}, 
  title={Detecting Anomaly in Big Data System Logs Using Convolutional Neural Network}, 
  year={2018},
  volume={},
  number={},
  pages={151-158},
  doi={10.1109/DASC/PiCom/DataCom/CyberSciTec.2018.00037}}

@inproceedings{suppa-etal-2021-cost,
    title = "Cost-effective Deployment of {BERT} Models in Serverless Environment",
    author = "Suppa, Marek  and
      Bene{\v{s}}ov{\'a}, Katar{\'i}na  and
      {\v{S}}vec, Andrej",
    editor = "Kim, Young-bum  and
      Li, Yunyao  and
      Rambow, Owen",
    booktitle = "Proceedings of the 2021 Conference of the North American Chapter of the Association for Computational Linguistics: Human Language Technologies: Industry Papers",
    month = jun,
    year = "2021",
    address = "Online",
    publisher = "Association for Computational Linguistics",
    url = "https://aclanthology.org/2021.naacl-industry.24/",
    doi = "10.18653/v1/2021.naacl-industry.24",
    pages = "187--195"
}

@INPROCEEDINGS{Sedki2023,
  author={Sedki, Issam and Hamou-Lhadj, Abdelwahab and Ait-Mohamed, Otmane and Ezzati-Jivan, Naser},
  booktitle={2023 IEEE/ACM 31st International Conference on Program Comprehension (ICPC)}, 
  title={Towards a Classification of Log Parsing Errors}, 
  year={2023},
  volume={},
  number={},
  pages={84-88},
  doi={10.1109/ICPC58990.2023.00023}}

@inproceedings{devlin-etal-2019-bert,
    title = "{BERT}: Pre-training of Deep Bidirectional Transformers for Language Understanding",
    author = "Devlin, Jacob  and
      Chang, Ming-Wei  and
      Lee, Kenton  and
      Toutanova, Kristina",
    editor = "Burstein, Jill  and
      Doran, Christy  and
      Solorio, Thamar",
    booktitle = "Proceedings of the 2019 Conference of the North {A}merican Chapter of the Association for Computational Linguistics: Human Language Technologies, Volume 1 (Long and Short Papers)",
    month = jun,
    year = "2019",
    address = "Minneapolis, Minnesota",
    publisher = "Association for Computational Linguistics",
    url = "https://aclanthology.org/N19-1423/",
    doi = "10.18653/v1/N19-1423",
    pages = "4171--4186"
}

@misc{distilbert,
      title={DistilBERT, a distilled version of BERT: smaller, faster, cheaper and lighter}, 
      author={Victor Sanh and Lysandre Debut and Julien Chaumond and Thomas Wolf},
      year={2020},
      eprint={1910.01108},
      archivePrefix={arXiv},
      primaryClass={cs.CL},
      url={https://arxiv.org/abs/1910.01108}, 
}

@misc{peronto2024logtrends,
  author       = {Riley Peronto},
  title        = {The State of Log Data: 6 Trends Impacting Observability and Security},
  howpublished = {Blog post, Chronosphere},
  year         = {2024},
  month        = {October},
  day          = {15},
  url          = {https://chronosphere.io/learn/observability-log-data-trends/}
}

@INPROCEEDINGS{Drain,
  author={He, Pinjia and Zhu, Jieming and Zheng, Zibin and Lyu, Michael R.},
  booktitle={2017 IEEE International Conference on Web Services (ICWS)}, 
  title={Drain: An Online Log Parsing Approach with Fixed Depth Tree}, 
  year={2017},
  volume={},
  number={},
  pages={33-40},
  doi={10.1109/ICWS.2017.13}}

@InProceedings{static_quantizatio,
  title = 	 {Up or Down? {A}daptive Rounding for Post-Training Quantization},
  author =       {Nagel, Markus and Amjad, Rana Ali and Van Baalen, Mart and Louizos, Christos and Blankevoort, Tijmen},
  booktitle = 	 {Proceedings of the 37th International Conference on Machine Learning},
  pages = 	 {7197--7206},
  year = 	 {2020},
  editor = 	 {III, Hal Daumé and Singh, Aarti},
  volume = 	 {119},
  series = 	 {Proceedings of Machine Learning Research},
  month = 	 {13--18 Jul},
  publisher =    {PMLR},
  url = 	 {https://proceedings.mlr.press/v119/nagel20a.html}
}

@INPROCEEDINGS{TinyLog,
  author={Meng, Chuangying and Chen, Ningjiang},
  booktitle={2024 International Joint Conference on Neural Networks (IJCNN)}, 
  title={TinyLog: Log Anomaly Detection with Lightweight Temporal Convolutional Network for Edge Device}, 
  year={2024},
  volume={},
  number={},
  pages={1-8},
  doi={10.1109/IJCNN60899.2024.10651312}}

@article{LightLog,
    title = {LightLog: A lightweight temporal convolutional network for log anomaly detection on the edge},
    journal = {Computer Networks},
    volume = {203},
    pages = {108616},
    year = {2022},
    issn = {1389-1286},
    doi = {https://doi.org/10.1016/j.comnet.2021.108616},
    url = {https://www.sciencedirect.com/science/article/pii/S1389128621005119},
    author = {Zumin Wang and Jiyu Tian and Hui Fang and Liming Chen and Jing Qin}
}

@INPROCEEDINGS {LogTransfer,
author = { Chen, Rui and Zhang, Shenglin and Li, Dongwen and Zhang, Yuzhe and Guo, Fangrui and Meng, Weibin and Pei, Dan and Zhang, Yuzhi and Chen, Xu and Liu, Yuqing },
booktitle = { 2020 IEEE 31st International Symposium on Software Reliability Engineering (ISSRE) },
title = {{ LogTransfer: Cross-System Log Anomaly Detection for Software Systems with Transfer Learning }},
year = {2020},
volume = {},
ISSN = {},
pages = {37-47},
doi = {10.1109/ISSRE5003.2020.00013},
url = {https://doi.ieeecomputersociety.org/10.1109/ISSRE5003.2020.00013},
publisher = {IEEE Computer Society},
address = {Los Alamitos, CA, USA},
month =Oct}

@article{LogPal,
author = {Sun, Lei and Xu, Xiaolong},
title = {LogPal: A Generic Anomaly Detection Scheme of Heterogeneous Logs for Network Systems},
journal = {Security and Communication Networks},
volume = {2023},
number = {1},
pages = {2803139},
doi = {https://doi.org/10.1155/2023/2803139},
url = {https://onlinelibrary.wiley.com/doi/abs/10.1155/2023/2803139},
eprint = {https://onlinelibrary.wiley.com/doi/pdf/10.1155/2023/2803139},
year = {2023}
}

@INPROCEEDINGS{EdgeLog,
  author={Chen, Jining and Chong, Weitu and Yu, Siyu and Xu, Zhun and Tan, Chaohong and Chen, Ningjiang},
  booktitle={2022 Tenth International Conference on Advanced Cloud and Big Data (CBD)}, 
  title={TCN-based Lightweight Log Anomaly Detection in Cloud-edge Collaborative Environment}, 
  year={2022},
  volume={},
  number={},
  pages={13-18},
  doi={10.1109/CBD58033.2022.00012}}

@article{zhang2020qbert,
  title={Ternarybert: Distillation-aware ultra-low bit bert},
  author={Zhang, Wei and Hou, Lu and Yin, Yichun and Shang, Lifeng and Chen, Xiao and Jiang, Xin and Liu, Qun},
  journal={arXiv preprint arXiv:2009.12812},
  year={2020}
}

@misc{Word2Vec,
      title={Integrating Distributional Lexical Contrast into Word Embeddings for Antonym-Synonym Distinction}, 
      author={Kim Anh Nguyen and Sabine Schulte im Walde and Ngoc Thang Vu},
      year={2016},
      eprint={1605.07766},
      archivePrefix={arXiv},
      primaryClass={cs.CL},
      url={https://arxiv.org/abs/1605.07766}, 
}

@article{Fasttext,
  author       = {Armand Joulin and
                  Edouard Grave and
                  Piotr Bojanowski and
                  Matthijs Douze and
                  Herv{\'{e}} J{\'{e}}gou and
                  Tom{\'{a}}s Mikolov},
  title        = {FastText.zip: Compressing text classification models},
  journal      = {CoRR},
  volume       = {abs/1612.03651},
  year         = {2016},
  url          = {http://arxiv.org/abs/1612.03651},
  eprinttype    = {arXiv},
  eprint       = {1612.03651},
  bibsource    = {dblp computer science bibliography, https://dblp.org}
}

@article{RT-Log,
author = {Jia, Peng and Cai, Shaofeng and Ooi, Beng Chin and Wang, Pinghui and Xiong, Yiyuan},
title = {Robust and Transferable Log-based Anomaly Detection},
year = {2023},
issue_date = {May 2023},
publisher = {Association for Computing Machinery},
address = {New York, NY, USA},
volume = {1},
number = {1},
url = {https://doi.org/10.1145/3588918},
doi = {10.1145/3588918},
journal = {Proc. ACM Manag. Data},
month = may,
articleno = {64},
numpages = {26}
}

@misc{hespeler2025,
      title={Temporal cross-validation impacts multivariate time series subsequence anomaly detection evaluation}, 
      author={Steven C. Hespeler and Pablo Moriano and Mingyan Li and Samuel C. Hollifield},
      year={2025},
      eprint={2506.12183},
      archivePrefix={arXiv},
      primaryClass={stat.ML},
      url={https://arxiv.org/abs/2506.12183}, 
}

@misc{USENIX_CFDR,
  title        = {The Computer Failure Data Repository (CFDR)},
  howpublished = {\url{https://www.usenix.org/cfdr}},
  note         = {Accessed: 2025-09-08},
  author       = {{USENIX Association}},
}

@misc{onnx-intro,
  title        = {ONNX: Open Neural Network Exchange — Introduction},
  author       = {{ONNX Project}},
  howpublished = {\url{https://onnx.ai/onnx/intro/}},
  note         = {Accessed: 2025-09-11},
  year         = {2025}
}

@misc{bert2018,
  author = {{Google Research}},
  title = {{BERT: Pre-training of Deep Bidirectional Transformers for Language Understanding}},
  howpublished = {\url{https://github.com/google-research/bert}},
  year = {2018},
  note = {Accessed: 2024-03-14}
}
